\documentclass{ar2e}
\usepackage{ulem,url,graphicx,longtable}  
\def\simge{\mathrel{\rlap{\raise 0.511ex
     \hbox{$>$}}{\lower 0.511ex \hbox{$\sim$}}}}
\def\simle{\mathrel{\rlap{\raise 0.511ex
      \hbox{$<$}}{\lower 0.511ex \hbox{$\sim$}}}}
\begin{document}

\input psfig.sty

\jname{Annu. Rev. Nucl. Part. Phys.}
\jyear{2012}
\jvol{62}
\ARinfo{1056-8700/97/0610-00}

\title{The Nuclear Equation of State and Neutron Star Masses}

\markboth{LATTIMER}{NUCLEAR EOS AND NEUTRON STAR MASSES}

\author{James M. Lattimer
\affiliation{Stony Brook University, Dept. of Physics \& Astronomy, Stony Brook, New York 11794-3800;\\ e-mail: lattimer@astro.sunysb.edu}}

\begin{keywords}
neutron stars, dense matter equation of state, nuclear symmetry energy 
\end{keywords}

\begin{abstract}
  Neutron stars are valuable laboratories for the study of dense
  matter.  Recent observations have uncovered both massive and
  low-mass neutron stars and have also set constraints on
  neutron star radii.  The largest mass measurements are powerfully influencing
  the high-density equation of state because of the existence of the
  neutron star maximum mass.  The smallest mass measurements, and the
  distributions of masses, have implications for the progenitors and
  formation mechanisms of neutron stars.  The ensemble of mass and
  radius observations can realistically restrict the properties of
  dense matter, and, in particular, the behavior of the nuclear
  symmetry energy near the nuclear saturation density.
  Simultaneously, various nuclear experiments are progressively
  restricting the ranges of parameters describing the symmetry
  properties of the nuclear equation of state.  In addition, new
  theoretical studies of pure neutron matter are providing insights.  These
  observational, experimental and theoretical constraints of dense
  matter, when combined, are now revealing a remarkable convergence.
\end{abstract}

\maketitle

\section{INTRODUCTION}

Neutron stars contain matter with the highest densities in the
observable universe.  As such, they are valuable laboratories for the
study of dense matter.  They are so compact that general relativity is
essential to their structures; indeed, the existence of a maximum
neutron star mass is a manifestation of general relativity.  Their
compactness, however, makes them challenging to study.  The vast
majority of neutron stars are observed as pulsars \cite{Manchester05},
and nearly 2000 have been detected (see {\tt http://www.atnf.csiro.au/research/pulsar/psrcat} and {\tt http://www.mpifr-bonn.mpg.de/oldpifr/div/pulsar/data/archive.html}
for pulsar databases).  Currently, however, only a few
aspects of neutron stars can be inferred from pulsar observations, such as
masses, spin rates, rough ages, and magnetic field strengths, although
outstanding accuracy in mass measurements ($<0.1\%$) is sometimes
achieved \cite{Weisberg10,Kramer06}.

Other crucial properties that are important for understanding the structure and
evolution of neutron stars, such as radii and surface temperatures,
can be estimated in a few cases from optical and X-ray observations of
cooling neutron stars as well as from X-ray bursts and flares
occurring on neutron star surfaces (see Section \ref{sec:radius}).  One
newly formed neutron star was observed in neutrinos\cite{Bionta87} in
the aftermath of the supernova SN 1987A, but only the roughest estimates for its
mass, binding energy and radius were revealed \cite{LY89}.  A galactic
supernova today would result in the detection of perhaps hundreds of
times more neutrinos, which should allow for more precise estimates.  In
the near future, observations of gravitational waves containing mass
and radius information \cite{Bauswein12} from merging
binaries containing neutron stars are expected.  Unfortunately, at
present, precision measurements of both the mass and the
radius of individual neutron stars do not exist. Current goals are to
combine available observations to deduce the underlying equation of
state (EOS) of dense matter, the subject of this review, and to infer
their internal composition from their cooling histories
\cite{Lattimer94,Page09}.

At the same time, ongoing laboratory measurements of nuclear
properties and theoretical studies of pure neutron matter are limiting
the properties of neutron star matter near the saturation density
(usually expressed as baryon density, $n_s\simeq0.16$ baryons
fm$^{-3}$; mass density, $\rho_s\simeq2.7\times10^{14}$ g cm$^{-3}$; or
energy density, $\varepsilon_s\simeq150$ MeV fm$^{-3}$).  In this
review, we show that estimates of the EOS from astrophysical observations are
converging with those from laboratory studies.  This development is important, because it means that the range of EOS properties that
need to be explored during simulations of supernovae and neutron
star mergers is shrinking.

In Section \ref{sec:structure}, we present the basics of neutron star structure, and we explore the limits with the aid of the 
  maximally compact EOS.  In Section \ref{sec:mass}, we present techniques for
extracting the masses of neutron stars in binary systems, and tabulate the mass measurements.  We summarize the results for
the maximum and minimum neutron star masses and the distribution of
neutron star masses.  We describe other observations in which
simultaneous mass and radius information can be obtained 
in Section \ref{sec:radius} and discuss the computation of the universal mass-radius
and pressure-density relations from observations in
Section \ref{sec:m-r}.  Finally, in Section \ref{sec:lab} we summarize the available laboratory nuclear physics constraints, as
well as recent theoretical studies of pure neutron matter.

\section{NEUTRON STAR STRUCTURE AND THE MAXIMALLY COMPACT EQUATION OF STATE\label{sec:structure}}

\subsection{The Mass-Radius Relation\label{sec:relation}}

\begin{figure}[h]
\vspace*{-.5cm}
\hspace*{-2.5cm}\includegraphics[width=6.5in,angle=180]{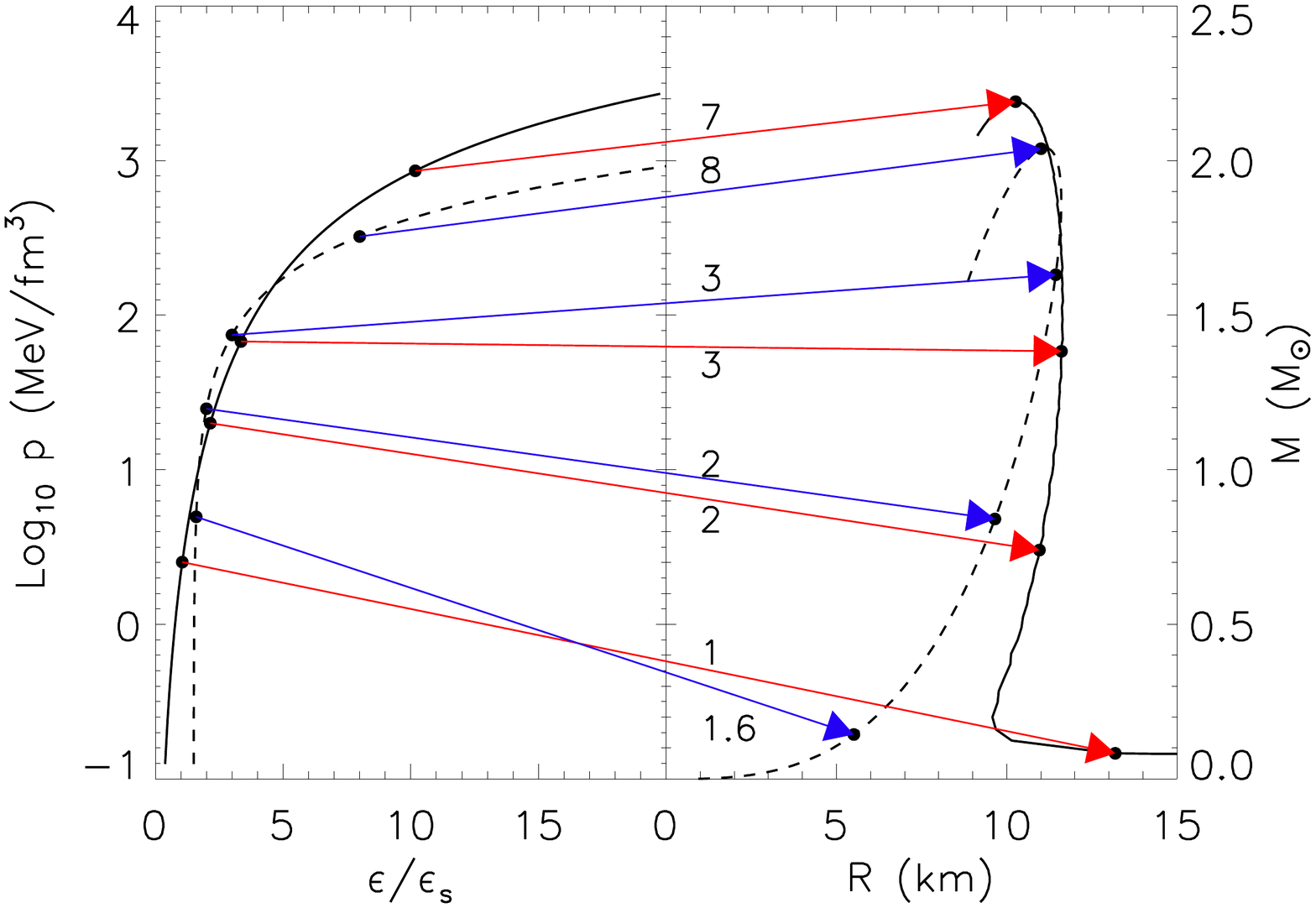}
\vspace*{0pt}
\caption{(a) Schematic hadronic (solid curve) and pure strange quark matter
  (dashed curve) equations of state.  (b) The corresponding $M-R$ relations.
  Arrows connect specific central energy density and pressure values with
  their corresponding $(M,R)$ points.  The numbers labelling hadronic arrows
  denote central baryon densities ($n_c/n_s$) and those labelling strange
quark matter 
  arrows indicate ($\varepsilon_c/\varepsilon_s$).  The upper-most
  arrows in each case mark the maximum mass configurations.}
\label{eosmr}\end{figure}

At very low temperatures, and for matter in weak-interaction
equilibrium, the only aspect of the EOS that is relevant for the
global structure of neutron stars is the pressure-energy density
relation.  The composition of matter, usually parameterized by the
number of electrons per baryon $Y_e$, is implicitly determined at each
density $n$ by minimizing the energy density $\varepsilon$, resulting in
the condition of $\beta$ equilibrium,
\begin{equation}\label{beta} \left({\partial\varepsilon/n\over\partial
    Y_e}\right)_n=\mu_e+\mu_p-\mu_n=0,
\end{equation} 
where the chemical potentials of neutrons, protons and electrons are
$\mu_n$, $\mu_p$ and $\mu_e$, respectively.  In this case, the general relativistic equations of hydrostatic
equilibrium \cite{OV39} are expressed by
\begin{eqnarray}
{dp\over dr}&=&-{G\over c^2}{(p+\varepsilon)(m+4\pi r^3p/c^2)\over r(r-2Gm/c^2)},\nonumber\\
{dm\over dr}&=&4\pi r^2{\varepsilon\over c^2},
\label{TOV}
\end{eqnarray}
where $p$ is the pressure, $m$ is the enclosed mass and $r$ is the radius.
Beginning from the center where $m=r=0$, $p=p_c$ and
$\varepsilon=\varepsilon_c$, these equations are integrated to the
surface, where $p=p_{surf}=0$, $r=R$, and $m=M$.  For hadronic EOSs,
$\varepsilon_{surf}=0$, but this is not the case for pure strange
quark matter (SQM) stars, for which the energy density remains finite.
Figure~\ref{eosmr} shows these features for schematic
hadronic and pure SQM EOSs.
The hadronic EOS consists of two polytropes (i.e., $p=Kn^\gamma$)
with exponents $\gamma_1=4/3$ and $\gamma_2=3$ joined at the
transition pressure $p_t=p_s/8$ where $p_s=2.5$ MeV fm$^{-3}$.  The
transition occurs at baryon density $n_s/2$.  The SQM EOS is the
simple MIT bag model $\varepsilon=4B+3p$ \cite{Chodos74} where the bag
constant is chosen to be $B=3\varepsilon_s/8$.

In the Newtonian limit, the equations of hydrostatic equilibrium are $dp/dr=-G\rho m/r^2$ and $dm/dr=4\pi\rho r^2$, where $\rho=nm_b$.  Dimensional analysis 
for a polytropic EOS results in scalings between the 
radius $R$ and total mass $M$:
\begin{equation}\label{scaling}
  R\propto K^{1/(3\gamma-4)}M^{(\gamma-2)/(3\gamma-4)}, \qquad
  M\propto K^{1/(2-\gamma)}R^{(3\gamma-4)/(\gamma-2)}.
\end{equation}
For the case in which $\gamma\sim4/3$, valid for hadronic matter at densities below
$\rho_s/3$ where the pressure is dominated by relativistic degenerate electrons, $M\propto K^{3/2}R^0$ which is
independent of radius.  At extremely large radii ($R\simge300$ km),
the mass starts to increase as configurations approach the white dwarf
(WD) range (such configurations have much larger proton concentrations).  Thus, there is
a {\it minimum stable mass} for neutron stars, which is approximately 0.09
M$_\odot$ \cite{HZD02}, when $R\sim200-300$ km.  

\begin{figure}[h]
\vspace*{-.5cm}
\includegraphics[width=5in]{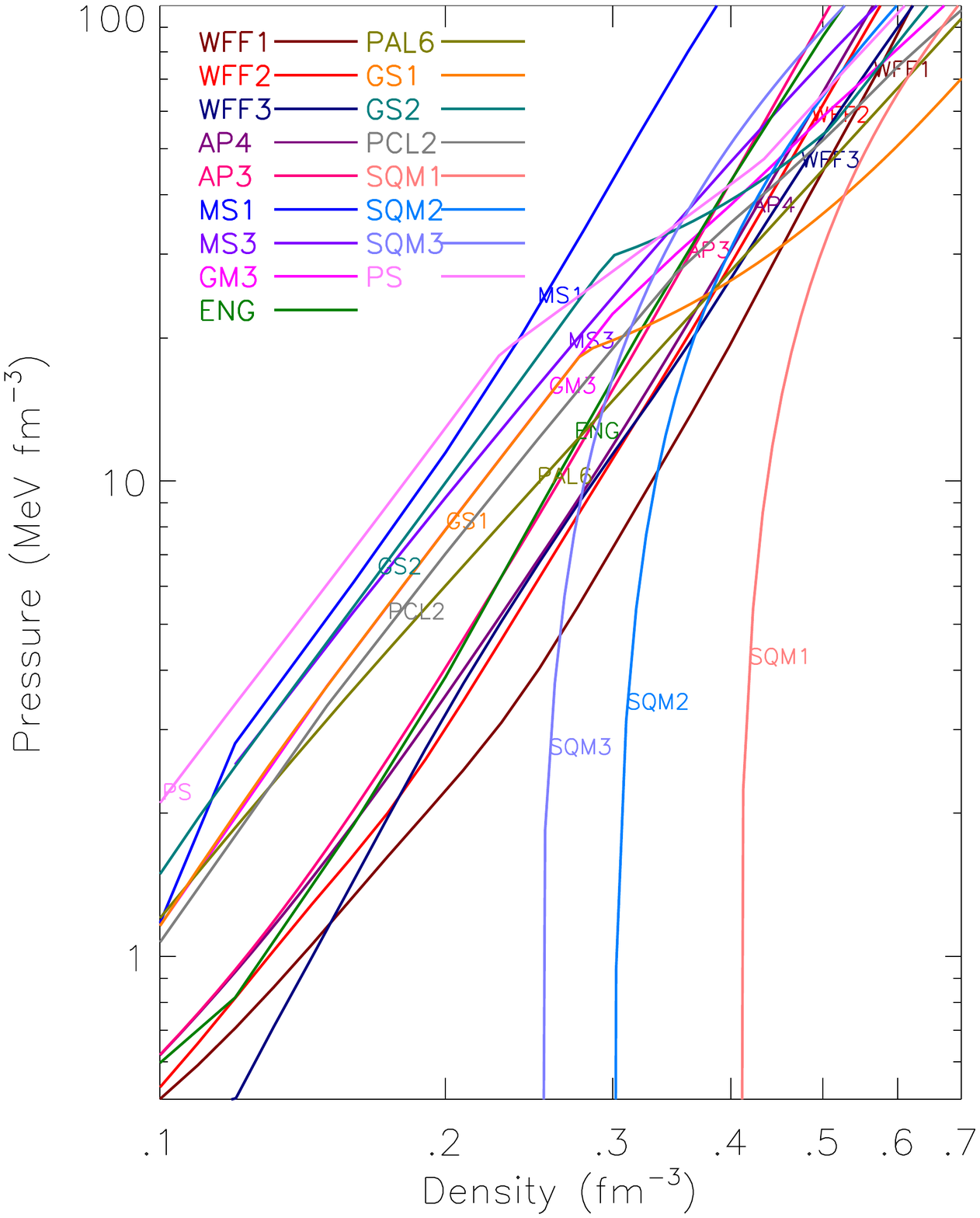}\vspace*{0pt}
\caption{Representative hadronic and strange quark matter (SQM) equations of state.  The mean exponent $\gamma=d\ln p/d\ln n\simeq2$ holds for hadronic EOSs in the vicinity of $n_s=0.16$ fm$^{-3}$. The range of pressures at $n_s$ is approximately a factor of six.  This figure is taken from and the EOS names are identified in Reference~\cite{LP01}.}
\label{prho}\end{figure}

For higher densities, in the range $\varepsilon_s-3\varepsilon_s$, the
typical behavior of hadronic EOSs is $\gamma\sim2$ (Figure~\ref{prho}).  In this case, the scaling becomes $R\propto
K^{1/2}M^0$, and the radii are nearly independent of mass.  

In the hadronic cas,e both asymptotic behaviors are apparent in
Figure~\ref{eosmr}, and the transition between them occurs
near $n_c\sim n_s$.  At high densities, general relativity becomes
dominant and causes the formation of a maximum mass.  In the case
of SQM stars at low densities, the large value of $B$ essentially results in
$\gamma\rightarrow\infty$ so that $R\propto K^0M^{1/3}$, a behavior
also shown in Figure~\ref{eosmr}.

An interesting feature of Figure~\ref{eosmr} is that the SQM and hadronic
EOSs predict very similar $M-R$ trajectories in the
range $1.5<M/{\rm M}_\odot<2$.  It would clearly be difficult on the
basis of observational $M-R$ data alone to distinguish these trajectories. This
observation suggests that other data, such as neutron star cooling information \cite{Page00},
will be necessary to confirm the existence of SQM stars.

\begin{figure}[h]
\vspace*{-.5cm}
\hspace*{-4cm}\includegraphics[width=7.5in,angle=180]{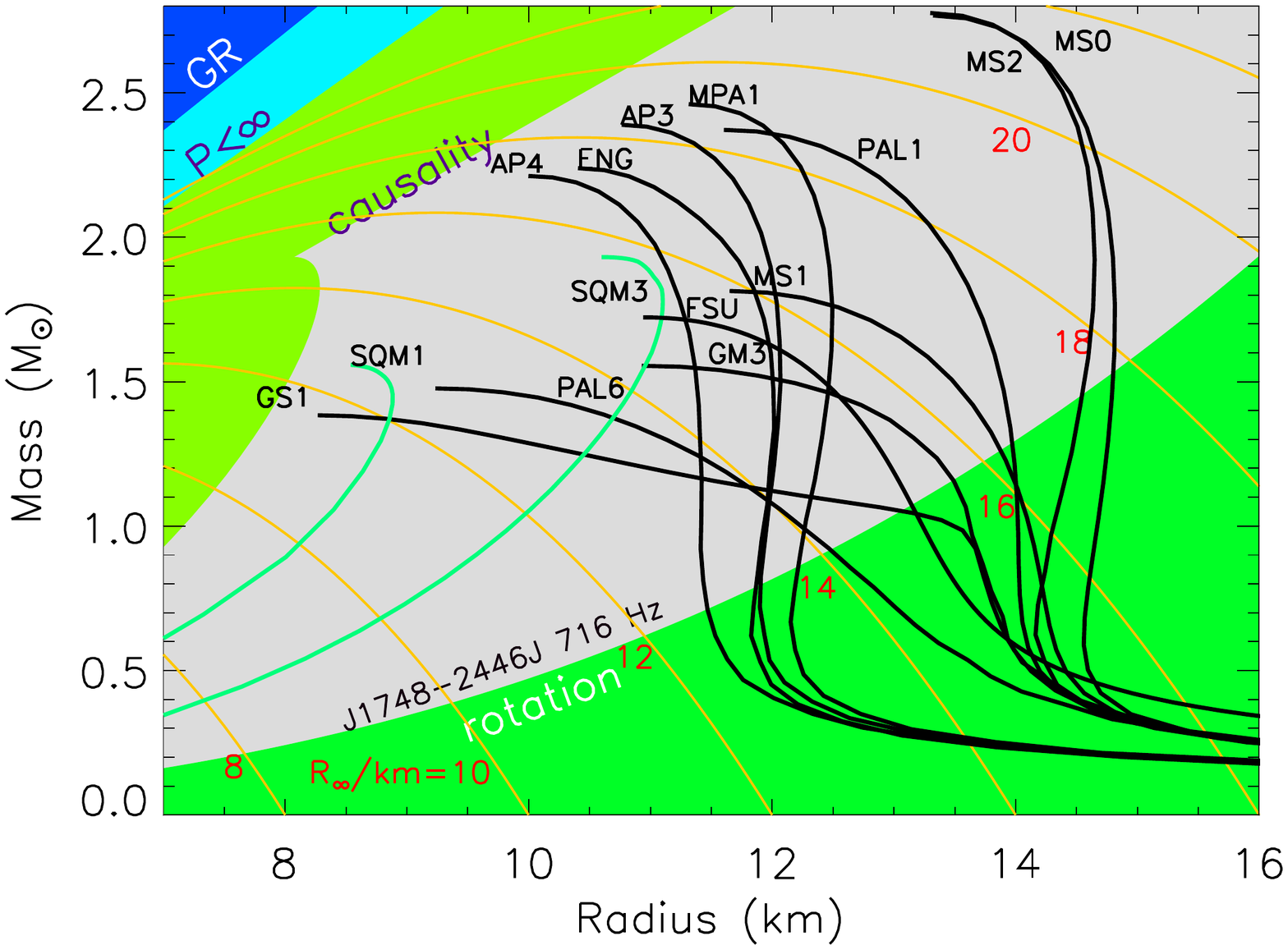}
\vspace*{0pt}
\caption{Typical $M-R$ curves for hadronic (black curves) and SQM
  (green curves) equations of state (EOSs).  The EOS names are identified in
  Reference~\cite{LP01} and their $P-n$ relations are displayed in Figure~\ref{prho}.
  Regions of the $M-R$ plane excluded by general relativity, finite pressure, and
  causality are indicated.  Orange curves show contours of
  $R_\infty=R(1-2GM/Rc^2)^{-1/2}$.  The region marked ``rotation''
  is bounded by the realistic mass-shedding limit for the highest known pulsar frequency, 716 Hz, for PSR J1748-2446J \cite{Hessels06}.  Figure adapted from Reference~\cite{LP07}.}
\label{mr}\end{figure}

It is useful to display $M-R$ curves for various realistic EOSs, as
demonstrated in Figure~\ref{mr} for several of the EOSs plotted in
Figure~\ref{prho}.  Those hadronic EOSs with extreme softening (due to a
kaon or pion condensate, high abundances of hyperons, or a low-density
quark-hadron phase transition) do not have pronounced vertical
segments, but they also do not allow the existence of a 2 M$_\odot$
neutron star (see Section 3) and, therefore, cannot be physical.  The $M-R$
curves that have attained sufficient mass have vertical segments with
radii varying from 10 to 16 km (the extreme cases are not shown in
Figure~\ref{mr}).  However, in contrast to the expectation that
$R\propto K^{1/2}$ when $\gamma\sim2$, it has been shown
phenomenologically \cite{LP01} that the scaling is approximately
$R\propto p_s^{1/4}$; the reduced exponent is due to general
relativistic effects.  Note that the ordering of radii for 1.4
M$_\odot$ neutron stars in Figure~\ref{mr} is commensurate with the
ordering of $p_s$ values in Figure~\ref{prho}.

In contrast, at low central densities where the pressure is dominated by degenerate
relativistic electrons and $\gamma\sim4/3$, we should expect
$M$ at a given $R$ to scale as $M\propto K^{3/2}$.  Indeed, two families
of hadronic stars are evident (EOSs AP4, ENG, AP3 and MPA1 constitute
one family in Figure~\ref{mr}), due to each family's relative
values of $p_s$ (Figure~\ref{prho}). 
  Thus, $M$ for a given $R$ on the
horizontal part, and $R$ for a given $M$ on the vertical part, 
increase with $p_s$.  

The pressure in the vicinity of $n_s$ for neutron
star matter is an important property of the nuclear EOS.  It is traditional to express the energy per baryon of hadronic matter near $n_s$ as a double Taylor expansion in $n-n_s$ and neutron excess $1-2x$, where $x$ is the proton fraction:
\begin{equation}
e(u,x)=-B+{K_o\over18}(u-1)^2+{K_o^\prime\over162}(u-1)^3 +S_2(u)(1-2x)^2+e_\ell+\dots.
\label{eose}\end{equation}
Here, $B\simeq16$ MeV is the bulk binding energy of symmetric matter at $n_s$, $K_o$ and $K_o^\prime$ are the standard incompressibility and skewness parameters, respectively; $S_2$ is the symmetry energy to quadratic order in $1-2x$; $e_\ell$ is the lepton energy; and $u=n/n_s$.   For bulk matter in $\beta$ equilibrium, when $u\simge0.01$ and $x<<1$, the lepton contributions are small.     If higher-order rather than quadratic terms in the neutron excess are unimportant, $S_2(u)\simeq S(u)$, where $S(u)$, the total symmetry energy, is the difference between pure neutron ($x=0$) and symmetric ($x=1/2$) matter energies.  The pressure is
\begin{equation}
p(u,x)=u^2n_s\left({\partial e\over\partial u}\right)_x\simeq u^2n_s\left[{K_o\over9}(u-1)+{K_o^\prime\over54}(u-1)^2+{dS_2\over du}(1-2x)^2\right]+p_\ell+\dots,
\label{eosp}\end{equation}
where $p_\ell$ is the lepton pressure.  In the vicinity of $u\simeq1$,
with $x<<1$, $p_\ell$ is small and the pressure is almost completely
determined by $dS/du$.
Laboratory constraints on the nuclear symmetry energy are discussed in
Section \ref{sec:lab}.

\subsection{The Maximally Compact EOS}
 Koranda et al.~\cite{KSF97} suggested that absolute limits to neutron star
 structure could be found by considering a soft low-density EOS
 coupled with a stiff high-density EOS, which would maximize the
 compactness $M/R$.  The limiting case of a soft EOS is $p=0$.  The
 limiting case of a stiff EOS is $dp/d\varepsilon=(c_s/c)^2=1$, where
 $c_s$ is the adiabatic sound speed that should not exceed light
 speed; otherwise, causality would be violated.  The maximally
   compact EOS is therefore defined by
\begin{equation}
p=0\quad{\rm for~}\varepsilon<\varepsilon_0;\qquad p=\varepsilon-\varepsilon_0\quad{\rm for~}\varepsilon>\varepsilon_0.
\label{mc}
\end{equation}
This EOS has a single parameter, $\varepsilon_0$, and therefore the structure equations (Equation~\ref{TOV}) can be expressed in a scale-free way:
\begin{equation}
{dw\over dx}=-{(y+4\pi x^3w)(2w-1)\over x(x-2y)},\qquad {dy\over dx}=4\pi x^2w,
\label{mctov}
\end{equation}
Here, $w=\varepsilon/\varepsilon_0$, $x=r\sqrt{G\varepsilon_0}/c^2$,
and $y=m\sqrt{G^3\varepsilon_0}/c^4$.  Varying the value of $w$ at the
origin ($w_0$) gives rise to a family of solutions described by dimensionless radius $X$ and total mass $Y$.  The maximally compact solution has the largest value of $Y$, which occurs when $w_0=3.034$, $X_c=0.2404$, and $Y_c=0.08513$.
Because $X_c/Y_c=2.824$ for this case, stable configurations satisfy
\begin{equation}
R\ge2.824{GM\over c^2},\qquad z=\left(1-{2GM\over Rc^2}\right)^{-1/2}-1\le0.8512,
\label{rgm}
\end{equation}
where $z$ is the redshift.  This limiting
redshift is close to the result obtained by Lindblom~\cite{Lindblom83}.  It is
also clear that the neutron star maximum mass is
\begin{equation}
M_{max}=Y_c{c^4\over\sqrt{G^3\varepsilon_0}}\simeq4.09\sqrt{\varepsilon_s\over\varepsilon_0}{\rm~M}_\odot.
\label{mmax}
\end{equation}
This result is similar to that found by Rhoades \& Ruffini~\cite{RR74}: If
the low-density EOS is known up to $\approx2\varepsilon_s$, causality
limits the neutron star maximum mass to approximately 3 M$_\odot$.

\begin{figure}[h]
\hspace*{-2.75cm}\includegraphics[width=7in,angle=180]{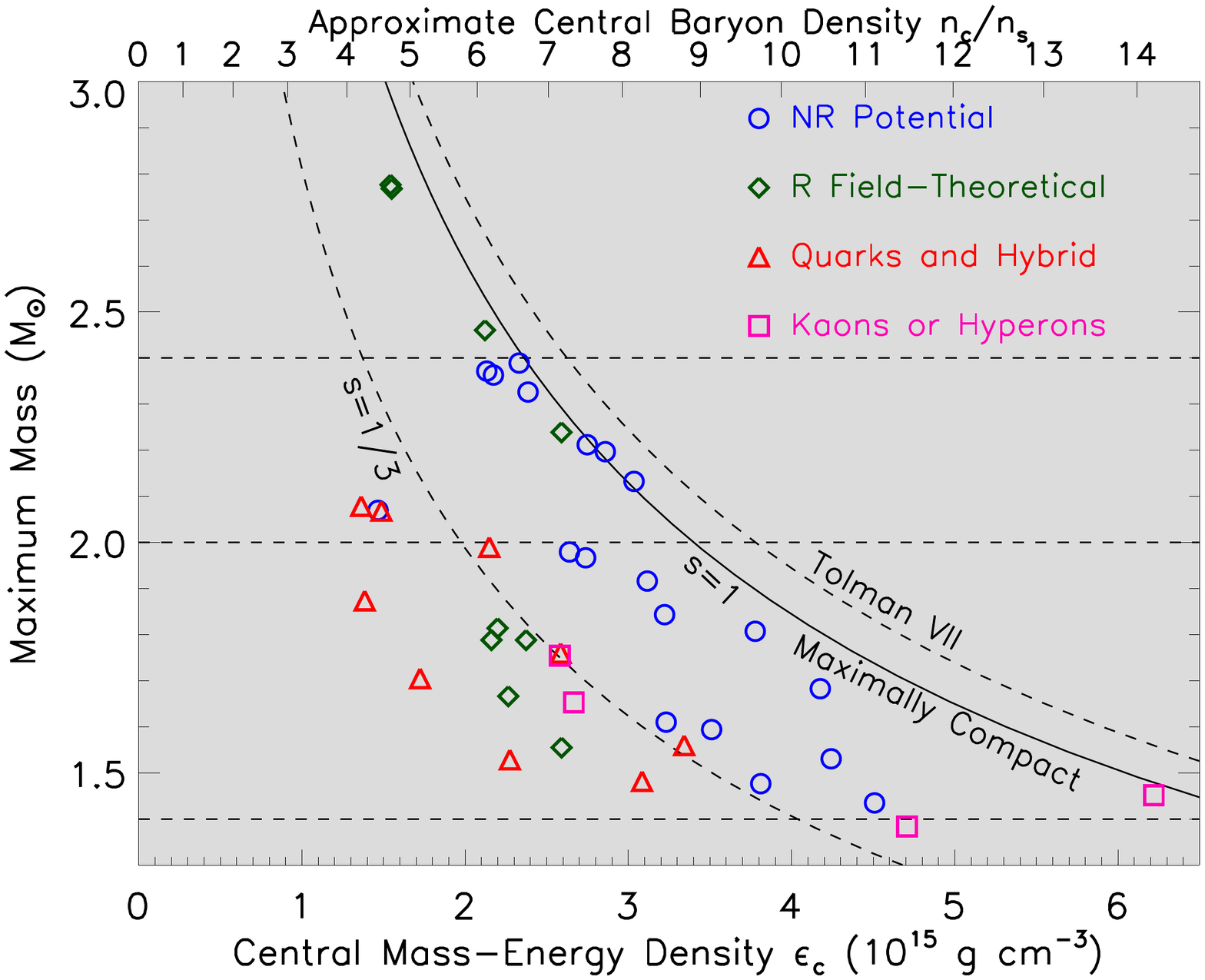}
\vspace*{0pt}
\caption{The relation between the maximum mass and the central density predicted by
  the maximally compact equation of state (EOS) (Equation~\ref{emax}) is shown by the
  curve $s=1$.  Results for different classes of EOSs are indicated:
  non-relativistic (NR) potential or Skyrme-like (open circles),
  relativistic (R) field theoretical (diamonds), EOSs containing
  significant amounts of quarks (triangles), and EOSs with
  significant meson condensates or hyperons (squares).  The
  $M_{max}-\varepsilon_c$ relations predicted by causality coupled with the Tolman
  VII solution \cite{Tolman39} and the EOS $p=(\varepsilon-\varepsilon_0)/3$ are also
  indicated.  The latter effectively bounds stars containing quarks.  Figure adapted from Reference \cite{LP05}.}
\label{maxdens}\end{figure}

The maximally compact solution also implies that
\begin{equation}
\varepsilon_c\le0.02199~\varepsilon_0Y_c^{-2}\simeq50.8~\varepsilon_s\left({{\rm M}_\odot\over M_{max}}\right)^2.
\label{emax}
\end{equation}
This relation (Figure~\ref{maxdens}) states that the largest
measured neutron star mass, given that it is lower than the true neutron star
maximum mass, must set an upper limit to the central density of any
neutron star \cite{LP05}, and by extension, that it also limits the central pressure
to values lower than 34.0 $\varepsilon_s({\rm M}_\odot/M_{max})^2$.

The scale-free character of the maximally compact solution can
be extended to axial symmetry in general relativity with no additional
parameters \cite{KSF97}, and the minimum spin period, limited by
mass- shedding at the equator, occurs for the maximum mass
configuration and scales witth $1/\sqrt{\varepsilon_0}$.  The result
can be expressed as
\begin{equation}\label{pmin}
P_{min}=0.74\left({{\rm M}_\odot\over M_{max}}\right)^{1/2}\left({R_{max}\over10{\rm~km}}\right)^{3/2}{\rm~ms}=0.20~\left({M_{max}\over{\rm M}_\odot}\right){\rm~ms},
\end{equation}
where $M_{max}$ and $R_{max}$ correspond to values for the
non-rotating star.  This equation represents the formal limit imposed by
causality on the spin period, but a more realistic limit for an arbitrary neutron star with non-rotating values $M$ and $R$, which has been
phenomenologically determined \cite{LP04}, replaces the coefficient 0.74 ms in Equation~(\ref{pmin}) with 0.96 ms, and $M_{max}$ and $R_{max}$ with
$M$ and $R$ (Figure~\ref{mr}).  Also, 
Lattimer \& Prakash \cite{LP11} have shown that the maximum fractional binding energy and the baryon chemical potential are 25.2\% and $\mu_{B,max}=2.093$ GeV, respectively, independent of  $\varepsilon_c$ and $M_{max}$.

The MIT bag model EOS can be
expressed as $p=(\varepsilon-\varepsilon_0)/3$, where
$\varepsilon_0=4B$, so the sound speed is $c_s=c/\sqrt{3}$
everywhere.  The solution of Equation~\ref{TOV} with this EOS may also be
expressed in a scale-free manner, with maximum mass eigenvalues
$w_0= 4.826$, $X_c=0.1910$, $Y_c=0.05169$, and
$X_c/Y_c=3.696$.  Thus, the $M_{max}-\varepsilon_{c}$ relation
mirrors that of Equation~\ref{emax} but is a factor $0.05169/0.08513=0.607$
lower (Figure~\ref{maxdens}).  Interestingly, this curve
apparently bounds the central densities of not only pure SQM stars but
also stars containing significant amounts of quarks in a pure or
mixed state with hadrons, even though the quark matter EOSs
used can be more complex than the 
MIT bag model.  This observation has important consequences for the quark-hadron
phase transition density if deconfined quarks appear in neutron stars,
as we discuss further below.

\begin{figure}[h]
\vspace*{-.5cm}
\hspace*{-2.5cm}\includegraphics[width=6.5in,angle=180]{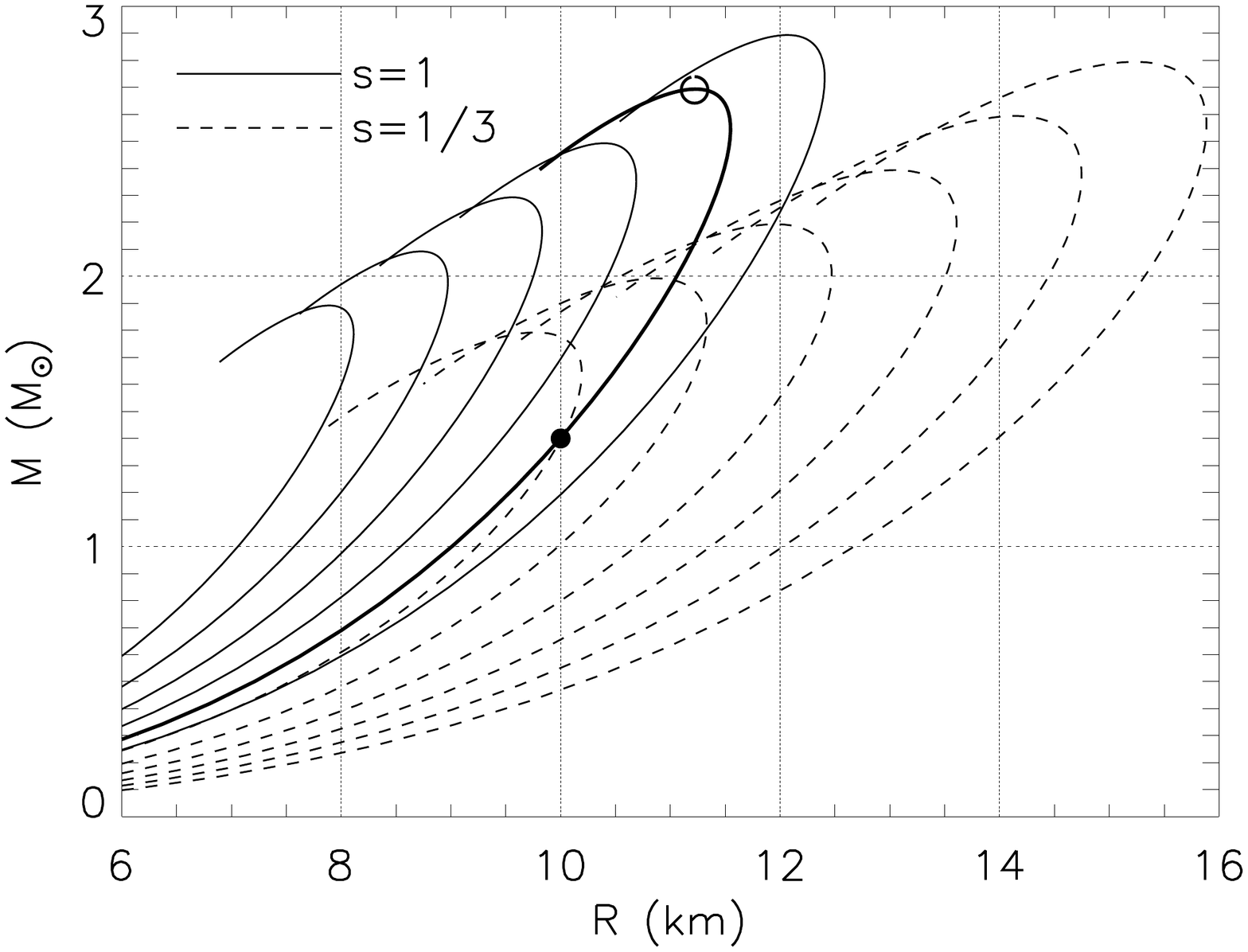}
\vspace*{0pt}
\caption{Mass-radius contours with various values of $M_{max}$ for the
  maximally compact equation of state (EOS) $p=\varepsilon-\varepsilon_0$ are shown as solid lines. Contours 
for the EOS $p=(\varepsilon-\varepsilon_0)/3$
  are shown as dashed lines.  A
  precise $(M,R)$ measurement (filled circle) sets an upper limit on $M_{max}$ (open circle) (2.69 M$_\odot$).}
\label{max}\end{figure}

Obviously, the largest precisely- known neutron star mass  
sets a lower bound to $M_{max}$.  A lower bound to $M_{max}$
also further restricts the allowed $M-R$ region beyond the area excluded by causality, $R\le2.824GM/c^2$ (Figure~\ref{mr}).  Conversely, the
precise measurements of any neutron star's mass and radius 
establishes an {\it upper} bound to $M_{max}$.
Figure~\ref{max} depicts these model-independent features and shows $M-R$ contours for the {\it
  maximally compact} EOS with various values of $\varepsilon_0$.  Each contour represents the minimum
radius possible for a given mass, given the contour's value of $M_{max}$.  For example, if $M_{max}\simge2$ M$_\odot$, a 1.4
M$_\odot$ neutron or quark star necessarily has $R\simge 8.25$ km.  It also follows
that any $M-R$ curve passing through a given ($M,R$) point
has a smaller value of $M_{max}$ than that of the {\it maximally
  compact} EOS passing through the same ($M,R$) point.  This hypothesis can be
easily demonstrated by overlaying $M-R$ trajectories for an alternate EOS,
such as the MIT bag-like model, $p=(\varepsilon-\varepsilon_0)/3$ (Figure~\ref{mr}).  For
the indicated point (1.4 M$_\odot$, 10 km), this alternate EOS has a
maximum mass of only 1.79 M$_\odot$ whereas the maximum mass of the the maximally compact
EOS is 2.69 M$_\odot$.

\section{NEUTRON STAR MASSES\label{sec:mass}}
\subsection{Mass Measurements}
The most accurate measurements concerning neutron stars are mass determinations from pulsar timing.  To date, approximately 33 relatively precise masses have become available.  In these systems, five Keplerian parameters can
be precisely measured \cite{MT77}; these parameters include the binary
period $P$, the projection of the pulsar's semimajor axis on the line
of sight $a_p\sin i$ (where $i$ is the orbit's inclination angle), the
eccentricity $e$, and the time $T_0$ and longitude $\omega$ of periastron.  Two of these observables yield a mass function:
\begin{equation}\label{fp} 
f_p=\left(\frac {2\pi}{P}\right)^2{\left(a_p\sin i\right)^3\over
  G}={(M_c\sin i)^3\over
  M^2}, 
\end{equation}
where $M=M_p+M_c$ is the total mass, $M_p$ is the pulsar mass, and
$M_c$ is the companion mass.  The minimum possible companion mass $M_c$ is equal to $f_p$.

The inclination angle $i$ is often the most difficult parameter to infer,
but even if it were known {\it a priori}, the above equation would 
specify a relation between only $M_p$ and $M_c$ unless the mass function
$f_c$ of the companion were also measurable.  The mass function of the companion is measurable in the rare case
when the companion itself is a detectable pulsar or a star with an observable
spectrum, as in an X-ray binary.
Fortunately, binary pulsars are compact systems, and general
relativistic effects can often be observed.  These effects include the advance
of the periastron of the orbit, 
\begin{equation}
\dot\omega=3\left(\frac {2\pi}{P}\right)^{5/3}\left({GM\over c^3}\right)^{2/3}
(1-e^2)^{-1};
\end{equation}
the
combined effect of variations in the tranverse Doppler shift and
gravitational redshift (time dilation) around an elliptical orbit,
\begin{equation}
\gamma=e\left(\frac{P}{2\pi}\right)^{1/3}\frac{M_c(M+M_c)}{M^{4/3}}
\left({G\over c^3}\right)^{2/3}; 
\end{equation} 
the orbital period decay due to the emission of gravitational
radiation, 
\begin{equation}
\dot P=-{192\pi\over5}\left({2\pi G\over c^3}\right)^{5/3}
\left(1+{73\over24}e^2+{37\over96}e^4\right)
(1-e^2)^{-7/2}{M_pM_c\over M^{1/3}};  
\end{equation} 
and Shapiro
time delay \cite{Shapiro64}, caused by the propagation of the pulsar signal
through the gravitational field of its companion.
The Shapiro delay produces a
delay in pulse arrival time \cite{Damour86},
\begin{equation} 
\delta_S(\phi)=2{GM_c\over c^3}\ln
\left[{1+e\cos\phi\over1-\sin(\omega+\phi)\sin i}\right],
\end{equation} 
where $\phi$ is the true anomaly, the angular parameter defining the
position of the pulsar in its orbit relative to the periastron.  $\delta_S$ 
is a periodic function of $\phi$ with an amplitude
\begin{equation}
\Delta_S=2{GM_c\over c^3}\left|\ln\left[\left({1+e\sin\omega\over1-e}\right)\left({1+\sin\omega\sin i\over1-\sin i}\right)\right]\right|,
\label{ds}
\end{equation} 
which is very large for extremely eccentric ($e\sim1$) and/or nearly
edge-on ($\sin i\sim1$) binaries.  For an edge-on circular orbit with $i\simeq\pi/2$,  
\begin{equation}\label{ds1}
\Delta_s\simeq4{GM_c\over c^3}\ln\left({2\over\cos i}\right)\,.
\end{equation}
\begin{figure}[h]
\vspace*{-.5cm}
\hspace*{-1.5cm}\includegraphics[width=6.5in,angle=180]{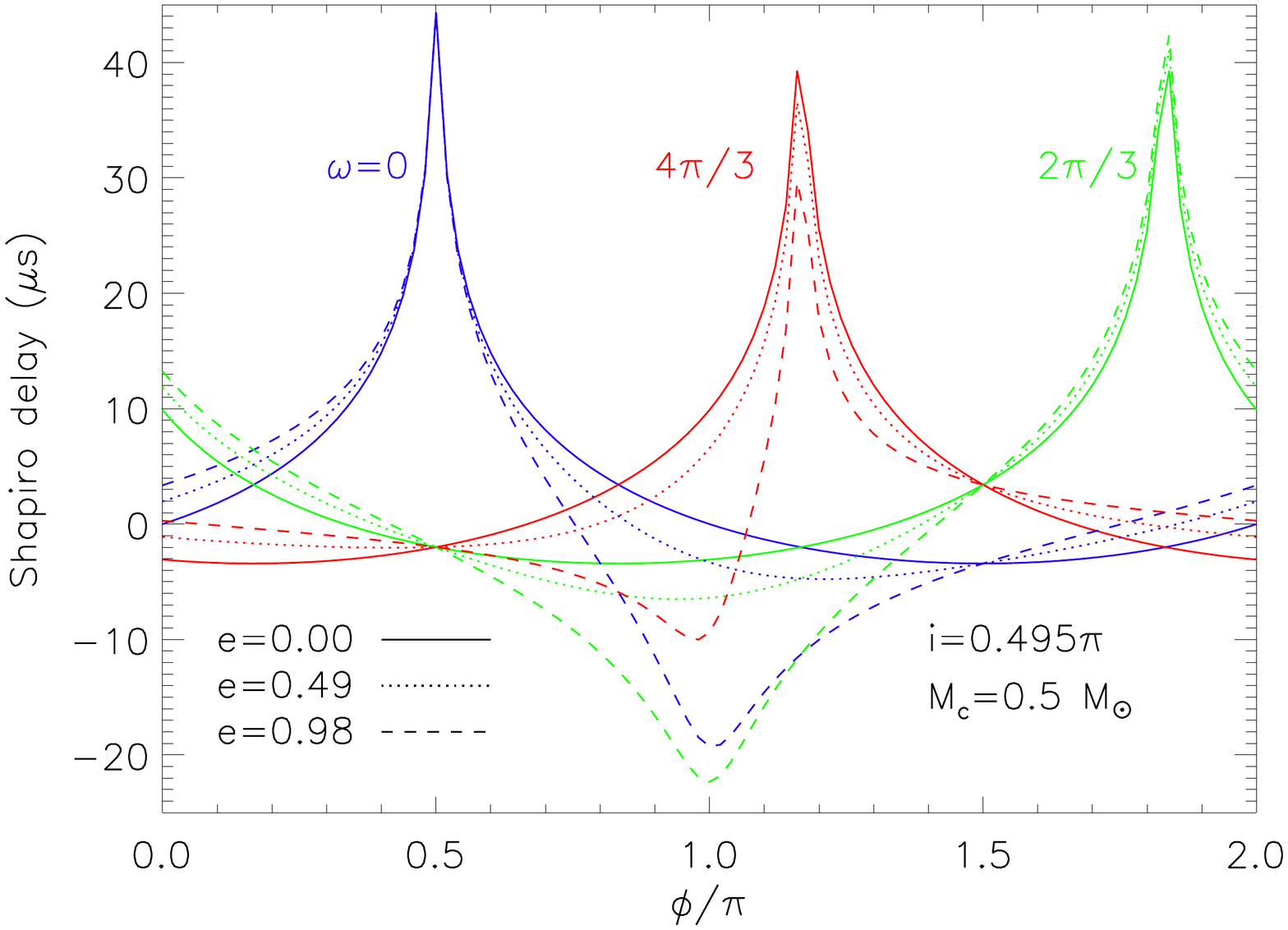}
\caption{Shapiro delay for various values of eccentricity $e$ and periastron longitude $\omega$ assuming an orbital inclination $i=0.498\pi$ radians.}
\label{shap}\end{figure}
The scale of Shapiro delay is set
by the constant $G{\rm M}_\odot/c^3=4.9255~\mu$s and both the
amplitude and shape of the Shapiro delay constrain $i, \omega$,
and $e$ (Figure ~\ref{shap}).

The inclination angle, and the individual masses, can be constrained
by measurements of two or more of these effects, which is possible
only in extremely compact systems.  Observation of a single
relativistic effect is not sufficient to constrain both masses.  Only
a small fraction of pulsars in binaries have two or more well-measured
relativistic effects that enable precise measurements of the pulsar
mass.  In some cases (Table~\ref{tab:mass}), an empirical relation
\cite{tauris99} between the binary period and the WD companion mass
proves useful.  In cases in which the companion is optically detected,
spectral observations can provide the WD mass and the radial velocity
amplitude, which can be used to establish the system's mass ratio.  In
some cases (e.g., References \cite{janssen08,gonzalez11}), the
inclination angle can be restricted by detection of changes in the
projected semi-major axis $x=a_p\sin i$ caused by the system's proper
motion \cite{Arzoumanian96},
\begin{equation}\label{proper}
\tan i=x\mu\sin\theta/(dx/dt),
\end{equation}
where $\theta$ is the unknown angle between the directions of the proper motion and the ascending node of the pulsar's orbit.
Use of $\sin\theta<1$ results in an upper limit to $i$ if $\mu, dx/dt$ and $x$ are measured, assuming it is known that $dx/dt$ is not due to gravitational radiation or perturbations caused by a third component.

In eclipsing X-ray binaries containing pulsars, the combination of radio timing data and X-ray observations yield the orbital period, eccentricity, longitude of periastron, the pulsar's orbital semi-major axis, and the eclipse duration.  Optical data can also provide the radial velocity amplitude of the companion and geometric information of the companion's shape.  WD
masses or surface gravities can be estimated from spectral measurements and effective
temperatures, if the distances are known.
Less accurate measurements of neutron star masses can be achieved for
X-ray binaries, in which X-rays are emitted by matter accreting onto a
neutron star from a companion star that is filling its Roche
lobe. Both X-ray and optical observations can yield both mass
functions, $f_p$ and $f_c$, but $f_c$ can be subject to large uncertainties
due to the faintness of optical radiation.  In some cases, the neutron star is a pulsar that establishes $f_p$.  Limits to inclination can
also be set according to the presence or absence of eclipses, given that
companions in X-ray binaries are not compact objects.

\begin{figure}
\hspace*{-1cm}\includegraphics[width=6in]{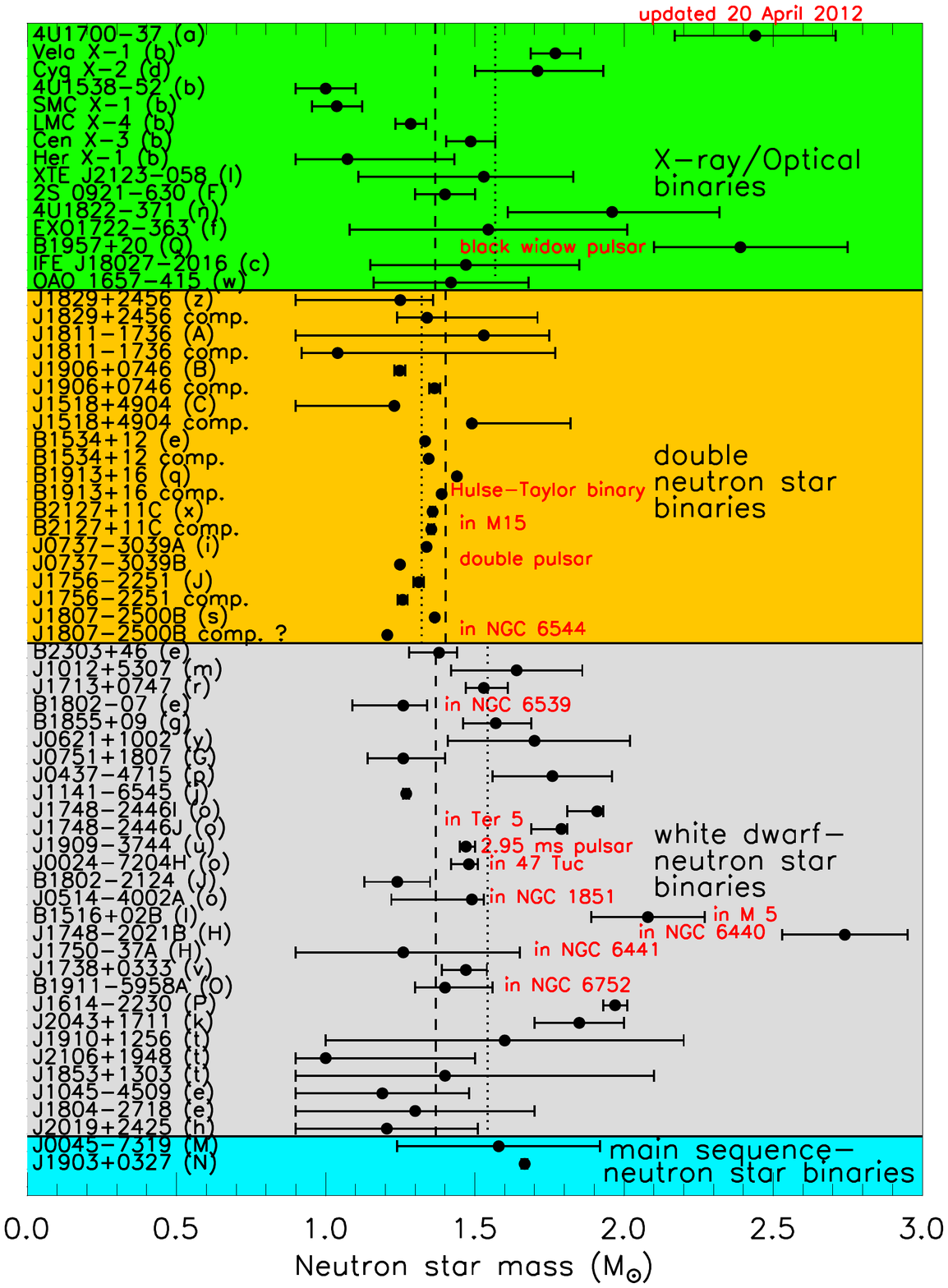}
\caption{Measured neutron star masses with 1-$\sigma$ errors. References in parenthesis following source names are identified in Table~\ref{tab:mass}.}
\label{masses}
\end{figure}

Figure~\ref{masses} and Table~\ref{tab:mass} summarize the available mass information from pulsars in binaries.  These summaries represent an update of those found in Reference~\cite{LP11}; I maintain a contemporary table, figure and references at \url{www.stellarcollapse.org}.  
Cases in which the standard millisecond pulsar formation model \cite{tauris99} has been used to constrain the neutron star mass are explicitly indicated in Table~\ref{tab:mass}.  Also, it has not yet been confirmed that the companion of PSR J1807-2500B \cite{lynch12} is a neutron star.  This system has the highest companion mass of any recycled pulsar.

\pagebreak
\begin{longtable}{ccc|ccc}
\caption{Neutron star mass measurements, 1-$\sigma$ uncertainties, $M>0.9$ M$_\odot$ assumed. 
}\label{tab:mass}\\

\hline Object & Mass (M$_\odot$) & \hspace*{-0.3cm}References &
Object & Mass (M$_\odot$) & \hspace*{-0.3cm}References \\
\endfirsthead
\multicolumn{6}{c}%
{{ \tablename\ \thetable{} --- continued from previous page}} \\

\hline Object & Mass (M$_\odot$) & \hspace*{-0.3cm}References & 
Object & Mass (M$_\odot$) & \hspace*{-0.3cm}References \\\hline\\[-3.5ex]
\endhead

\hline \multicolumn{6}{r}{{Continued on next page}} \\ 
\endfoot

\hline \hline
\endlastfoot
\hline\\[-6ex]
\multicolumn{6}{c}{\it X-ray/optical binaries (mean $1.568$ M$_\odot$, error-weighted mean $1.368$ M$_\odot$)}\\[-1ex]
4U 1700-377$^\dagger$ & $2.44^{+0.27}_{-0.27}$ & a \cite{clark02} &  
Vela X-1 & $1.770^{+0.083}_{-0.083}$ & b \cite{rawls11} \\[-2ex]

Cyg X-2 & $1.71^{+0.21}_{-0.21}$ & d \cite{casares10} & 
4U 1538-52 & $1.00^{+0.10}_{-0.10}$ & b \cite{rawls11} \\[-2ex]

SMC X-1 & $1.037^{+0.085}_{-0.085}$ & b \cite{rawls11} & 
LMC X-4 & $1.285^{+0.051}_{-0.051}$ & b \cite{rawls11} \\[-2ex]

Cen X-3 & $1.486^{+0.082}_{-0.082}$ & b \cite{rawls11} & 
Her X-1 & $1.073^{+0.358}_{-0.173}$ & b \cite{rawls11} \\[-2ex]

XTE J2123-058 & $1.53^{+0.30}_{-0.42}$ & l \cite{gelino03}; $\spadesuit$ & 
2S 0921-630 & $1.4^{+0.1}_{-0.1}$ & F \cite{steeghs07} \\[-2ex]

4U 1822-371 & $1.96^{+0.36}_{-0.35}$ & n \cite{munoz05} &
EXO 1722-363 & $1.545^{+0.465}_{-0.465}$ & f \cite{mason10}\\[-2ex]

B1957+20 & $2.39^{+0.36}_{-0.29}$ & Q \cite{vankerkwijk11} &
IGR J18027-2016 & $1.47^{+0.38}_{-0.32}$ & c \cite{mason11}\\[-2ex]

OAO 1657-415 & $1.42^{+0.26}_{-0.26}$ & w \cite{mason12} &&&\\
\hline \\[-6ex]

\multicolumn{6}{c}{\it neutron star -- neutron star binaries (mean $1.322$ M$_\odot$, error-weighted mean $1.402$ M$_\odot$)}\\[-1ex]

J1829+2456$^\ddagger$ & $1.25^{+0.11}_{-0.35}$ & z \cite{champion05} &
Companion & $1.34^{+0.37}_{-0.10}$ & z \cite{champion05} \\[-2ex]

J1811-1736$^\ddagger$ & $1.53^{+0.22}_{-0.63}$ & A \cite{corongiu07} &
Companion & $1.04^{+0.73}_{-0.12}$ & A \cite{corongiu07} \\[-2ex]

J1906+0746 & $1.248^{+0.018}_{-0.018}$ & B \cite{kasian08} &
Companion & $1.365^{+0.018}_{-0.018}$ & B \cite{kasian08} \\[-2ex]

J1518+4904 & $1.23^{+0.00}_{-0.33}$ & C \cite{janssen08} &
Companion & $1.49^{+0.33}_{-0.00}$ & C \cite{janssen08} \\[-2ex]

B1534+12 & $1.3332^{+0.0010}_{-0.0010}$ & K \cite{stairs02} & 
Companion & $1.3452^{+0.0010}_{-0.0010}$ & K \cite{stairs02} \\[-2ex]

B1913+16 & $1.4398^{+0.0002}_{-0.0002}$ & q \cite{weisberg10} &
Companion & $1.3886^{+0.0002}_{-0.0002}$ & q \cite{weisberg10} \\[-2ex]

B2127+11C$^\clubsuit$ & $1.358^{+0.010}_{-0.010}$ & x \cite{jacoby06} &
Companion$^\clubsuit$ & $1.354^{+0.010}_{-0.010}$ & x \cite{jacoby06} \\[-2ex]

J0737-3039A & $1.3381^{+0.0007}_{-0.0007}$ & i \cite{kramer06} & 
J0737-3039B & $1.2489^{+0.0007}_{-0.0007}$ & i \cite{kramer06} \\[-2ex]

J1756-2251 & $1.312^{+0.017}_{-0.017}$ & J \cite{ferdman08} &
Companion & $1.258^{+0.017}_{-0.017}$ & J \cite{ferdman08} \\[-2ex]

J1807-2500B$^\clubsuit$ & $1.3655^{+0.0020}_{-0.0020}$ & s \cite{lynch12} &
Companion ? & $1.2064^{+0.0020}_{-0.0020}$ & s \cite{lynch12} \\

\hline\\[-6ex]
\multicolumn{6}{c}{\it neutron star -- white dwarf binaries (mean $1.543$ M$_\odot$, error-weighted mean $1.369$ M$_\odot$)}\\[-1ex]

B2303+46 & $1.38^{+0.06}_{-0.10}$ & e \cite{thorsett99} & 
J1012+5307 & $1.64^{+0.22}_{-0.22}$ & m \cite{lange01} \\[-2ex]

J1713+0747$^*$ & $1.53^{+0.08}_{-0.06}$ & r \cite{splaver05} & 
B1802-07$^\clubsuit$ & $1.26^{+0.08}_{-0.17}$ & e \cite{thorsett99} \\[-2ex]

B1855+09$^*$ & $1.57^{+0.12}_{-0.11}$ & g \cite{nice03b} & 
J0621+1002 & $1.70^{+0.10}_{-0.17}$ & y \cite{nice08} \\[-2ex]

J0751+1807 & $1.26^{+0.14}_{-0.14}$ & y \cite{nice08} & 
J0437-4715 & $1.76^{+0.20}_{-0.20}$ & p \cite{verbiest08} \\[-2ex]

J1141-6545 & $1.27^{+0.01}_{-0.01}$ & j \cite{bhat08} & 
J1748-2446I$^\clubsuit$ & $1.91^{+0.02}_{-0.10}$ & o \cite{kiziltan11} \\[-2ex]

J1748-2446J$^\clubsuit$ & $1.79^{+0.02}_{-0.10}$ & o \cite{kiziltan11} &
J1909-3744$^\clubsuit$ & $1.47^{+0.03}_{-0.02}$ & u \cite{hotan06} \\[-2ex]

J0024-7204H$^\clubsuit$ & $1.48^{+0.03}_{-0.06}$ & o \cite{kiziltan11} &
B1802-2124 & $1.24^{+0.11}_{-0.11}$ & R \cite{ferdman10} \\[-2ex]

J0514-4002A$^\clubsuit$ & $ 1.49^{+0.04}_{-0.27}$ & o \cite{kiziltan11}  &
B1516+02B$^\clubsuit$ & $2.08^{+0.19}_{-0.19}$ & I \cite{freire08a} \\[-2ex]

J1748-2021B$^\clubsuit$ & $2.74^{+0.21}_{-0.21}$ & H \cite{freire08b}  &
J1750-37A$^\clubsuit$ & $1.26^{+0.39}_{-0.36}$ & H \cite{freire08b} \\[-2ex]

J1738+0333 & $1.47^{+0.07}_{-0.08}$ & v \cite{antoniadis12} & 
B1911-5958A$^\clubsuit$ & $1.34^{+0.08}_{-0.08}$ & O \cite{bassa06} \\[-2ex]

J1614-2230 & $1.97^{+0.04}_{-0.04}$ & P \cite{demorest10}  &
J2043+1711$^*$ & $1.85^{+0.15}_{-0.15}$ & k \cite{guillemot12} \\[-2ex]

J1910+1256$^*$ & $1.6^{+0.6}_{-0.6}$ & t \cite{gonzalez11} &
J2106+1948$^*$ & $1.0^{+0.5}_{-0.1}$ & t \cite{gonzalez11} \\[-2ex]

J1853+1303$^*$ & $1.4^{+0.7}_{-0.5}$ & t \cite{gonzalez11} &
J1045-4509 & $1.19^{+0.29}_{-0.29}$ & e \cite{thorsett99} \\[-2ex]

J1804-2718 & $1.3^{+0.4}_{-0.4}$ & e \cite{thorsett99} & 
J2019+2425 & $1.205^{+0.305}_{-0.305}$ & h \cite{nice01} \\
\hline \\[-6ex]

\multicolumn{6}{c}{\it NS -- Main Sequence Binaries}\\[-1ex]

J0045-7319 & $1.58^{+0.34}_{-0.34}$ & e \cite{thorsett99} & 
J1903+0327$^\heartsuit$ & $1.667^{+0.021}_{-0.021}$ & N \cite{freire11} \\
\end{longtable}
$^\dagger$Black hole due to lack of pulsations?

$^\ddagger$Companion masses from Reference \cite{thorsett99}

$^*$Binary period-WD masses from Reference \cite{tauris99}

$^\clubsuit$Globular cluster binary

$^\heartsuit3\sigma$ error

$\spadesuit$ J. Tomsick, private communication
\pagebreak

\subsection{The neutron star maximum mass}
From the perspective of EOS constraints, the most
massive neutron stars are key.  There is now ample observational
support from pulsars for neutron stars with masses significantly
greater than 1.5 M$_\odot$.  These include PSR J1903+0327
\cite{freire11}, which has a main-sequence companion; the globular cluster pulsars
PSR 1748-2446 I and J\cite{kiziltan11}, PSR J1748-2021B \cite{freire08a}, PSR B1516+02B
\cite{freire08b}, PSR J2043+1711 \cite{guillemot12}, and PSR J1614-2230 \cite{demorest10}, all of which have
WD companions; and the black widow pulsar B1957+20
\cite{vankerkwijk11}.  The inclination angles of the two binaries
containing pulsars I and J in the globular cluster Ter 5 are
unconstrained by observation, but if their inclinations are random, there is a 95\% chance that at least
one of these pulsars is greater than 1.68 M$_\odot$ \cite{ransom05}.
The WD binary period-mass relation \cite{tauris99} was used to restrict the inclination of PSR J2043+1711.  
Inclinations are also unknown for PSR J1748-2021B and PSR B1516+02B and are assumed to be random,
but this assumption is dangerous because the systems are not randomly selected.

The largest well-measured mass is $1.97\pm0.04$ M$_\odot$ 
for PSR J1614-2230 \cite{demorest10} for which $i$ has been determined by detection of Shapiro time delay.  
This 3.15 ms pulsar has an 8.7-day nearly
circular orbit with an 0.5-M$_\odot$ companion.  For this binary, the projected semimajor axis is $a_p\sin i=11.3$
light-seconds and $\sin i=0.99989$; in other words, it is virtually edge on.
The Shapiro time delay amplitude (from Equation {\ref{ds1}) is 48.8 $\mu$ s.
By virtue of its accuracy, this mass has become the standard for the minimum value of the
neutron star maximum mass.

In addition, a few X-ray binaries seem to contain high-mass
neutron stars: approximately 1.8 M$_\odot$ in the case of Vela X-1
\cite{rawls11}, 2 M$_\odot$ for 4U 1822-371 \cite{munoz05}, and 2.4
M$_\odot$ for 4U 1700-377 \cite{clark02} and PSR B1957+20
\cite{vankerkwijk11}.  Nonetheless, the large systematic errors
inherent in X-ray binary mass measurements warrant caution.

The case of the black widow pulsar, PSR B1957+20, represents an
intriguing case.  This system has both pulsar timing and
optical light curve information that yield both a mass function and an
estimate of the inclination $i$ from the shape of the light
curve \cite{Reynolds07}.  The binary consists of a 1.6-ms pulsar in a nearly
circular 9.17-h orbit around an extremely low mass companion:
$M_c\simeq0.03$ M$_\odot$.  The pulsar is eclipsed for approximately 10\% of
each orbit, but considering that $a_p\sin i=0.089$ light-second =
0.038 R$_\odot$ and that $a_c\sin i\sim3$ R$_\odot$ is $M_p/M_c\simeq80$
times larger, the eclipsing object has to be approximately $0.1a_c\sin
i\sim0.3$R$_\odot$---much larger than the size of the companion star.
Irradiation of the companion by the pulsar strongly heats its near side
to the point of ablation which leads to a comet-like tail and a large
cloud of plasma that is believed responsible for the eclipses.  The pulsar is
literally consuming its companion, hence the name black widow; it has
reduced its companion's mass to a small fraction of its original mass.  (In fact, several
black widow systems are known.)  The
irradiation also produces an enormous (factor--of--100) variation in the
brightness of the companion during its orbit.  The companion is
bloated and nearly fills its Roche lobe.  The companion's optical
light curve allows one to estimate the mass ratio $M_p/M_c$ and the inclination angle
$i$.  However, the large size of the companion means
that the ``center of light'' of the system is not equivalent to its
center of mass: The optical light curve depends on the projected
semi-major axis of the irradiated near side of the companion, rather
than the projected semi-major axis of the center of mass of the
companion.  The extreme case is either that the companion has zero
radius or that it completely fills its Roche lobe, leading to a range
$1.7<M_p/{\rm M}_\odot<3.2$, but estimates based on modeling have reduced
the probable range to $2.4\pm0.4$ M$_\odot$ \cite{vankerkwijk11}.  It will be valuable to
extend observations and modeling of this system because a 2.4 M$_\odot$
neutron star would have profound implications.

\subsection{The minimum neutron star mass}
In addition to large measured masses, there is increasing evidence for
small neutron stars.  Three X-ray binaries \cite{rawls11} and two neutron
stars in double neutron star binaries have best estimates of less than approximately
1.1 M$_\odot$, although in all cases the error bars are substantial.
The most interesting low-mass candidates are SMC X-1 and 4U 1538-52,
which have 1-$\sigma$ upper limits to mass of 1.122 M$_\odot$ and 1.10
M$_\odot$, respectively.  \"Ozel et al \cite{Ozel12} reanalyzed these systems without
the effects of the ellipsoidal shapes of the companion stars on the optical light curves,
and found masses of $0.93\pm0.12$ and  $1.18\pm0.25$, respectively.  According to either analysis, at least
one of these systems has a very low neutron star mass.   

Neutron star masses lower than 1.2 M$_\odot$ would challenge the
paradigm of gravitational-collapse neutron star formation.  The
iron cores of  $8-10$-M$_\odot$ progenitor stars are approximately the Chandrasekhar baryon mass, which corrected for electron
fractions $Y_e$ of less than 0.5 for finite temperatures, is approximately 1.25
M$_\odot$.  The lowest-mass neutron stars may form from
such pogenitor stars in so-called
electron-capture supernovae, in which an oxygen-neon-magnesium (O-Ne-Mg) core
collapses as pressure support is lost due to electron captures on Ne
and/or Mg nuclei \cite{Nomoto84}.  Electron captures are triggered by density increases due to accretion in binaries.  Correcting for
binding, the lowest-gravitational-mass neutron star could be 1.15-1.2 M$_\odot$.

A low-mass limit is also suggested by thermodynamics.
All gravitational-collapse supernova models, including O-Ne-Mg accretion-induced collapses, produce
hot, lepton-rich proto-neutron stars due to neutrino trapping during
infall.  Lepton fractions $Y_\ell=Y_e+Y_\nu$ (where $Y_\nu$ is the net
number of electron neutrinos per baryon) of order 0.3-0.35, and
entropies per baryon $s\sim1$, exist in the centers of neutron stars
at birth, according to current simulations.  Within 10 seconds, $Y_\nu$ within 
the star drops to zero through diffusion, but the bulk of the neutrino energy 
remains in the remnant as thermal energy \cite{Burrows86}.  Compared with cold,
catalyzed stars in $\beta$ equilibrium without neutrinos ($Y_\nu=0$),
which have a {\it minimum mass} of order 0.09 M$_\odot$ (as discussed
in Section~\ref{sec:relation}), proto-neutron stars have a
minimum gravitational mass of order 0.9-1.2 M$_\odot$, depending on
the entropy profile within the star \cite{Strobel99}.
Masses at the higher end of this range are suggested for
configurations with strongly shocked outer cores. Masses smaller than 
the minimum are dynamically unstable and cannot lead to stable neutron stars.

In summary, the current gravitational-collapse paradigm for cores of massive stars that lead to hot, lepton-rich proto-neutron
stars and successful supernova explosions imposes a lower limit on
neutron star masses at birth.  Neutron star masses can increase due to
fallback after the explosion and accretion from binary companions, but
they cannot further decrease except during a catastrophic merger.
Given the observational errors, the lowest observed masses are not currently incompatible with this
paradigm, but this remains an interesting problem.

\subsection{The distribution of neutron star masses}

Most neutron stars have masses close to 1.3 to 1.4 M$_\odot$, but lower
and higher masses exist.  Evolution probably plays many roles
in the distribution of neutron star masses: To name only two considerations, the neutron star birth
mass seems to depend on progenitor mass, and accretion can lead to the
accumulation of several tenths of a solar mass over a star's life.  The observed masses (Table~\ref{tab:mass}) may be
separated into four groups that could have different evolutionary
histories: X-ray binaries, double--neutron star binaries,WD--neutron star binaries, and WD--neutron star binaries
found in globular clusters.  Figire~\ref{histdist} shows histograms of
masses for these four groups, but they are of
limited utility given that individual stars can have
significant mass errors (1-$\sigma$ errors are given in Table~\ref{tab:mass}).
Interestingly, the first three groups in Table~\ref{tab:mass} have
the same error-weighted average masses to within 0.03 M$_\odot$.

\begin{figure}[h]
\hspace*{-1cm}\includegraphics[width=6in,angle=180]{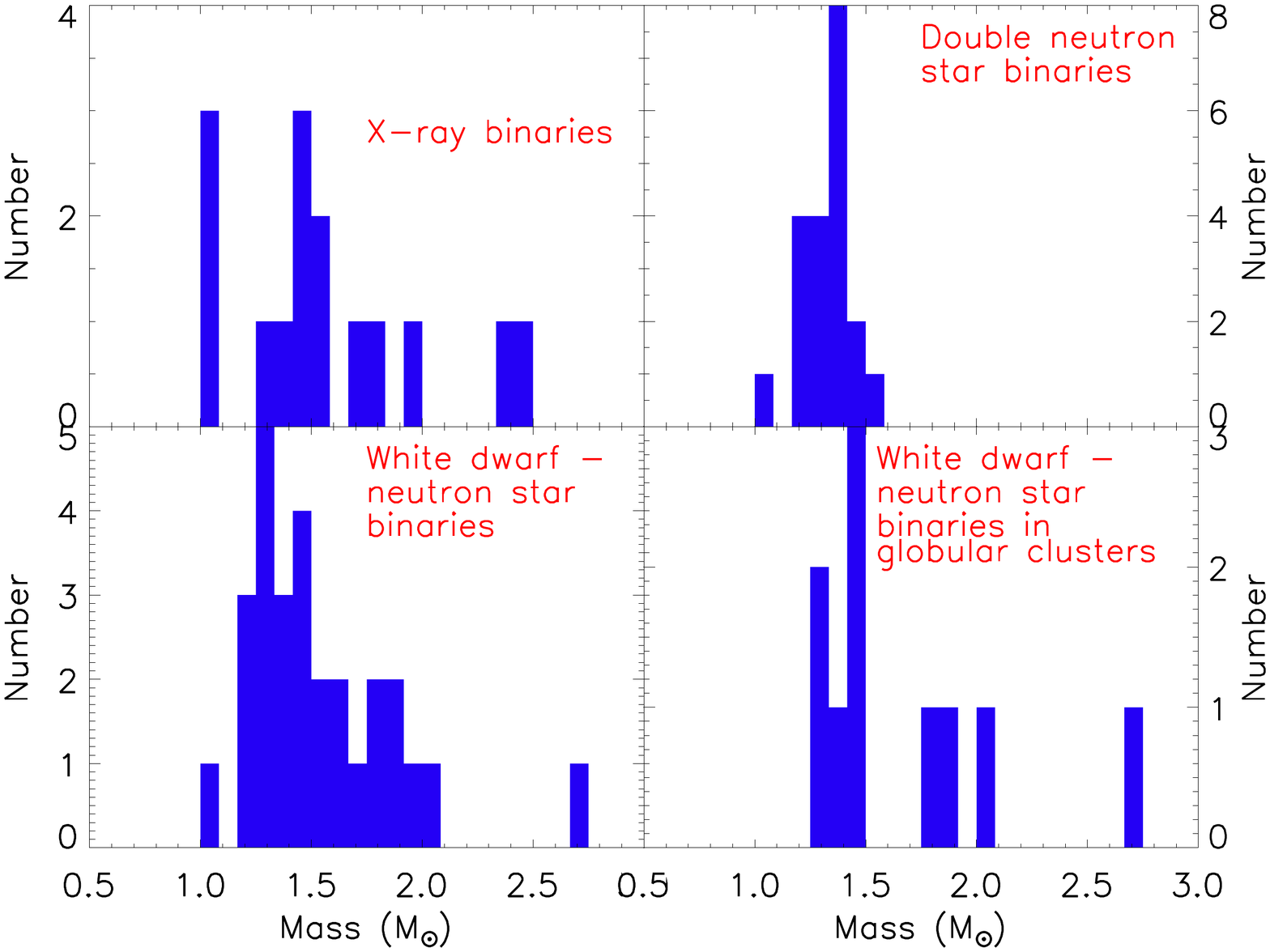}
\caption{Histograms of neutron star masses for four groups: X-ray
  binaries, double--neutron star binaries, white dwarf--pulsar
  binaries, and white dwarf--pulsar binaries in globular clusters.
  Bins are taken to be 0.08333 M$_\odot$ in width.}
\label{histdist}\end{figure}

To account for errors, one can approximate each star with
a Gaussian probability distribution whose integral over mass is unity.
The histograms are thereby transformed into a plot of the number of stars
in a small mass interval, or a density, as a function of mass. Those
stars with large errors have a relatively small contribution at any
mass; those with small errors contribute only over a very narrow
range of masses.  Figure~\ref{gaussdist} shows these densities.

\begin{figure}[h]
\hspace*{-1cm}\includegraphics[width=6in,angle=180]{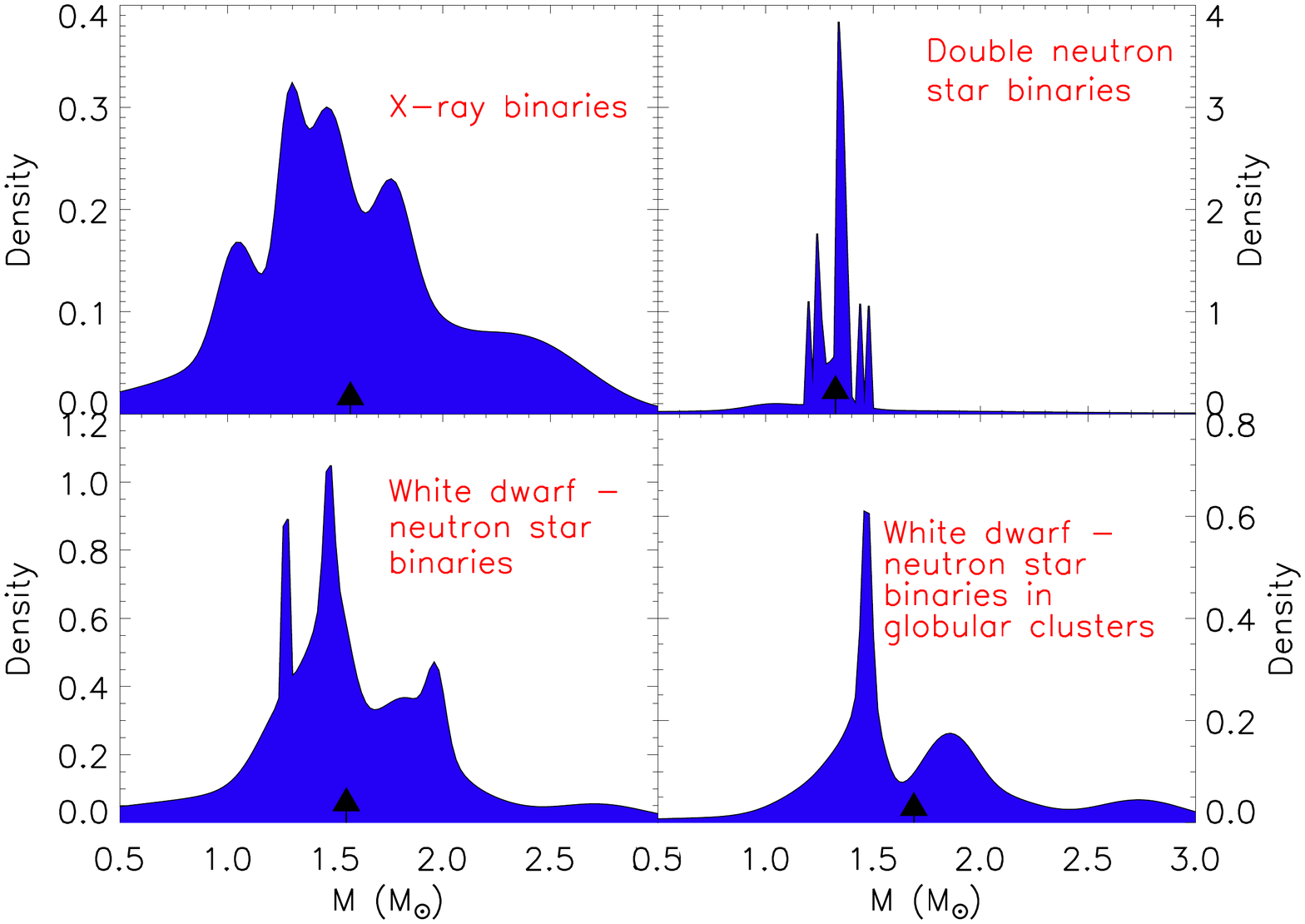}
\caption{The density of neutron star masses, namely, the number of stars per mass interval (in arbitrary units) is shown for the four groups of neutron stars displayed in Figure~\ref{histdist}.  The weighted average mass is indicated by the arrow in each plot.}
\label{gaussdist}\end{figure}

The most significant feature of the distribution of X-ray binaries is
a broad maximum.  However, the distribution of double--neutron stars is
very narrow and peaked at 1.33 M$_\odot$.  The distribution of
WD--neutron star binaries is more complex; there is evidence for three
groups with peaks at 1.25 M$_\odot$, 1.45 M$_\odot$ and $\sim1.9$
M$_\odot$.  The latter two peaks are also prominent in the
distribution of pulsar masses with WD compansions in globular
clusters.  Zhang et al.~\cite{Zhang11} noted the existence of the
first two groups.  They identified the first group with periods greater than 20 ms as non-recycled
pulsars, and the second group with smaller periods as
recycled pulsars, which have experienced extended mass-accretion episodes.  An analysis by \"Ozel et
al.~\cite{Ozel12} instead concludes there are three lower-mass groups, not
two: the first with a mean of 1.28 M$_\odot$ and dispersion of 0.24
M$_\odot$ in non-recycled high-mass binaries; the second
(representing stars in double--neutron star systems) with a mean of 1.33
M$_\odot$ and dispersion of 0.06 M$_\odot$; and a third containing recycled pulsars with a mean of
1.48 M$_\odot$.  

On the other hand, Schwab et al.~\cite{Schwab10}
argued for two peaks at low masses, 1.25 M$_\odot$ and 1.35 M$_\odot$.
The first peak represents neutron star production via O-Ne-Mg supernovae,
and the second peak represents neutron star production through conventional iron core
collapse.  The existence of two types of supernovae---iron
core-collapse supernovae from high-mass progenitors, and
electron-capture supernovae from low-mass O-Ne-Mg cores---is supported
by statistics of neutron star-hosting X-ray binaries.  Knigge et al~\cite{Knigge11} show that these binaries are composed
of two subpopulations differentiated by spin periods, orbital periods,
and orbital eccentricities.  Those with short spin and orbital periods
and low eccentricites probably originate from O-Ne-Mg
accretion-induced collapses.

The study by \"Ozel et al. demonstrates that the pulsar and companion
stars in double--neutron star binaries appear to be drawn from the
same distribution.  Their small dispersions could be attributed to
their birth in O-Ne-Mg supernovae, whose He cores are expected to have
a small mass range ($1.36-1.38$ M$_\odot$) prior to collapse
\cite{Podsiadlowski06}. \"Ozel et al.'s study also indicates that
although the mean mass of the lowest group of WD--meutron star
binaries is nearly the same as that of the double--neutron star
binaries, the dispersion of the former appears greater.  How
significant this finding is, however, depends on the details of
fitting the optical light curves in these systems.

In contrast, Kiziltan et al.~\cite{kiziltan11} claimed that there
is significant statistical evidence for only two groups centered at 1.35
M$_\odot$ and 1.5 M$_\odot$.  Presumably, only the
higher-mass group has experienced considerable accretion.

\section{SIMULTANEOUS MASS AND RADIUS MEASUEMENTS\label{sec:radius}}
In contrast to mass determinations, there are no high-accuracy radius measurements.  Moreover, there are no radius measurements for any neutron stars with a precise mass determination.  Many astrophysical observations that could lead to the extraction of neutron star radii, or combined mass and radius constraints, have been proposed.  These observations include the following.
\begin{enumerate}
\item Thermal X-ray and optical fluxes from isolated and quiescent neutron stars \cite{Mereghetti11};
\item Type I X-ray bursts on neutron star surfaces \cite{Lewin93};
\item Quasi-periodic oscillations from accreting neutron stars \cite{Miller06};
\item Spin-orbit coupling, observable through pulsar timing in extremely compact binaries, leading to moments of inertia \cite{LS05};
\item Pulsar glitches, which constrain properties of neutron star crusts \cite{Link99}; 
\item Cooling following accretion episodes in quiescent neutron stars that also constrain crusts \cite{Cackett06};
\item Neutron star seismology from X-rays observed from flares from soft $\gamma-$ray repeaters \cite{Samuelsson07};
\item Pulse profiles in X-ray pulsars, which constrain $M/R$ ratios due to gravitational light- bending \cite{Leahy09};
\item Gravitational radiation from tidal disruption of merging neutron stars \cite{Bauswein12}; 
\item Neutrino signals from proto-neutron stars formed in Galactic supernovae \cite{Burrows86}.
\end{enumerate}
Of these proposed observations, thermal emission and X-ray bursts from neutron star surfaces have dominated
recent attempts to infer neutron star radii.

\subsection{Thermal Emission from Quiescent and Isolated Sources}

Until a million years after a star's birth, neutrino emission dominates
thermal emission from the surface, but the star is observable
as an X-ray source (and, if near enough, as an optical source).
Several thermally- emitting neutron stars have been observed, some from 
nearby isolated sources and others from binaries in globular clusters.
To a zeroth approximation, thermal emission from neutron stars is blackbody, 
so measures of their integrated fluxes
and temperatures yield estimates of their angular diameters through
Kirchoff's laws.  However, because the observed flux is redshifted twice
and the temperature once, the inferred radius is not the geometric
radius $R$ but rather the so-called radiation radius:
$R_\infty=R/\sqrt{1-2GM/Rc^2}$.  A significant complication, however, is that the 
emissions are modulated by a star's atmosphere and magnetic
field; a neutron star is not a perfect blackbody. 

The principal uncertainties in extracting radii from thermal emissions include the following.
\begin{enumerate}
\item The distance $D$ (because the inferred $R_\infty\propto D$).
\item The magnitude of interstellar hydrogen (H) absorption, given that most hard UV radiation and an appreciable fraction of
X-rays are absorbed between the star and the Earth.
\item The atmospheric composition and magnetic field strength and distribution.
\end{enumerate}
The best chances of an accurate measurement are either from (a) 
 nearby isolated neutron stars for which parallax distances are
available (but have unknown atmospheric compositions and field
strengths), or (b) quiescent X-ray binaries in globular clusters with reliable
  distances.  Due to recent accretion episodes, these sources are
  believed to have low magnetic fields and, almost certainly,
  H-dominated atmospheres, which are the most reliably modeled \cite{Rutledge99}.

\begin{figure}
\begin{picture}(6,12)
\put(0,-60){\includegraphics[width=.45\columnwidth]{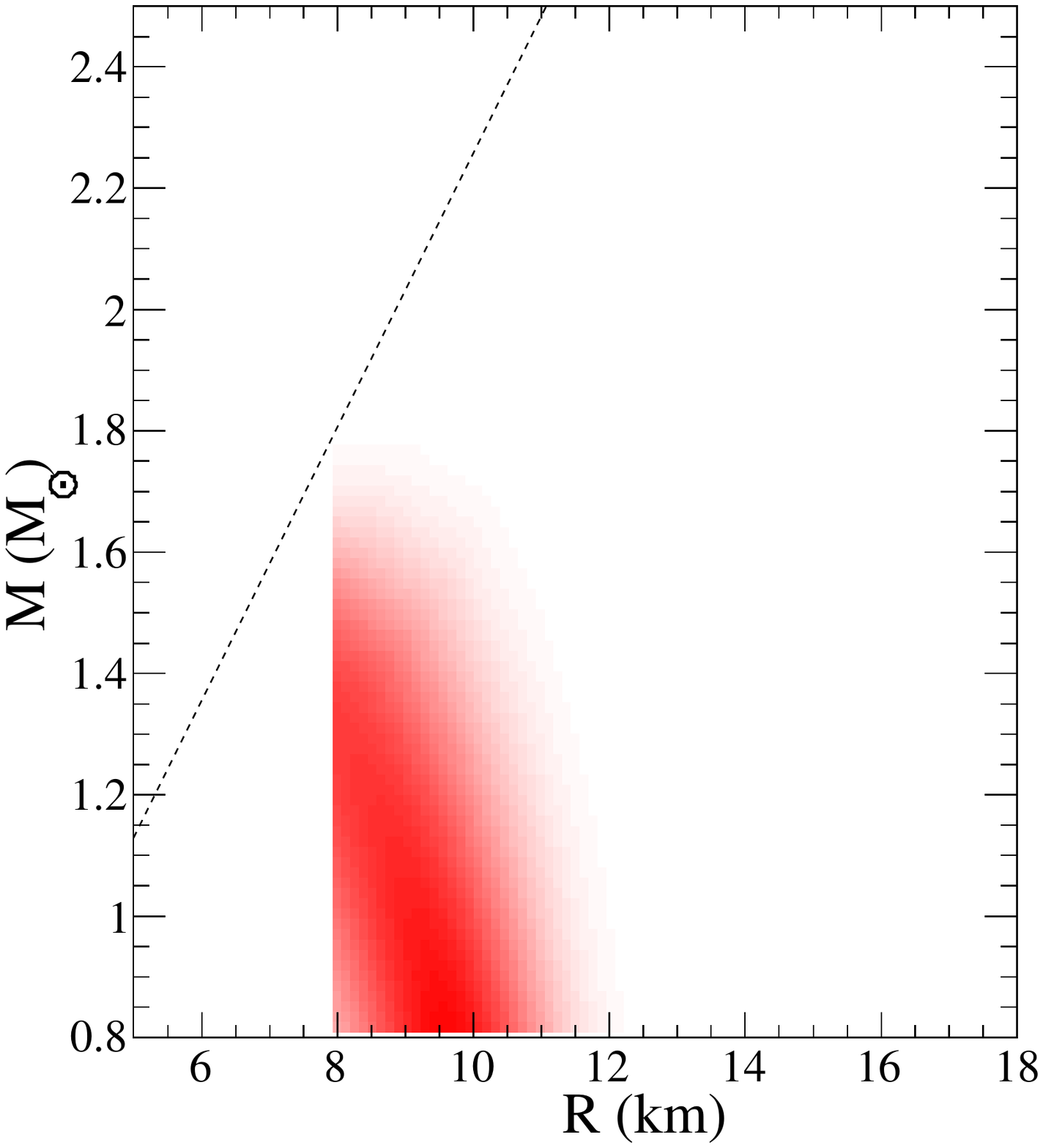}}
\put(140,-60){\includegraphics[width=.45\columnwidth]{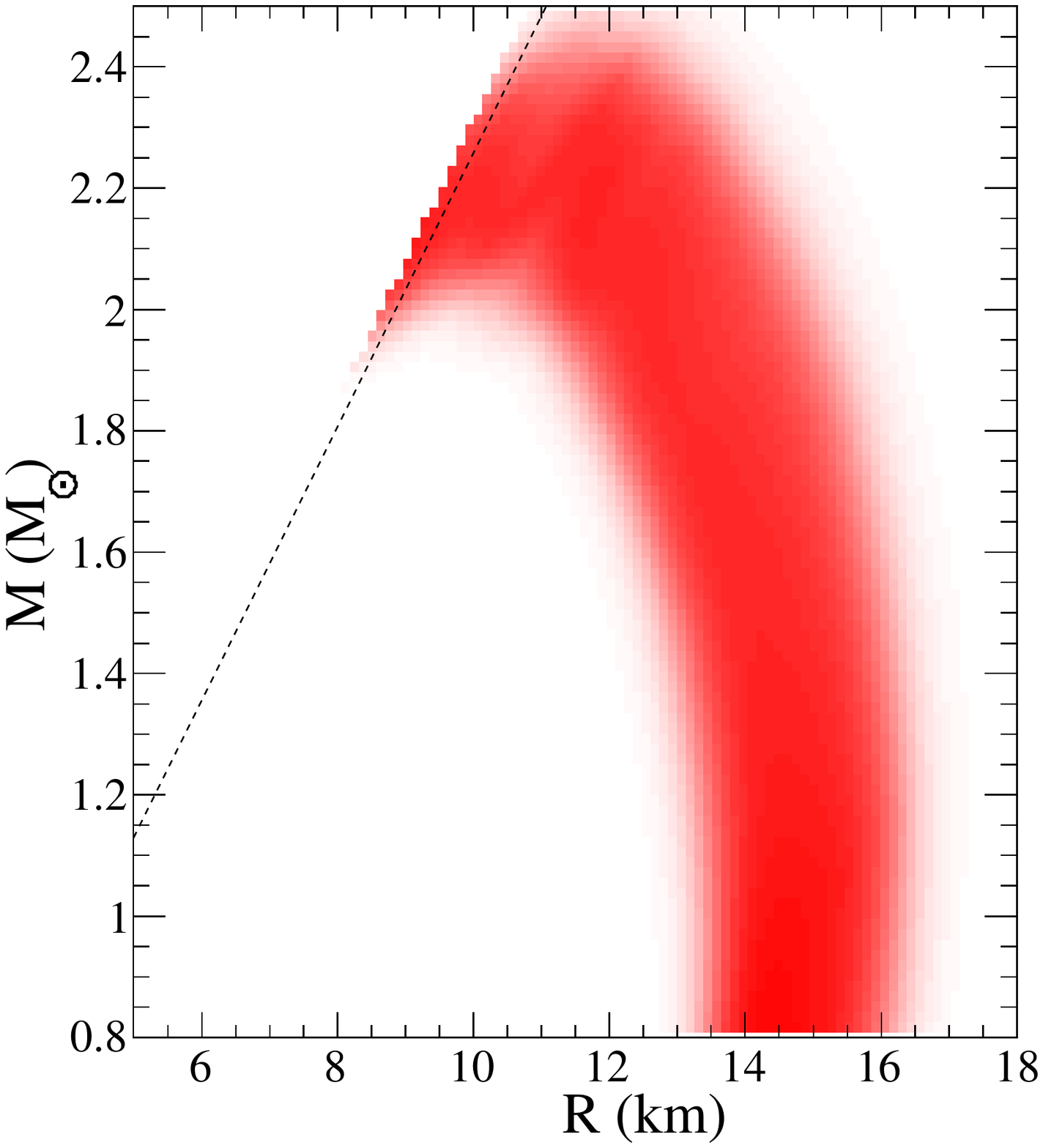}}
\put(280,-60){\includegraphics[width=.45\columnwidth]{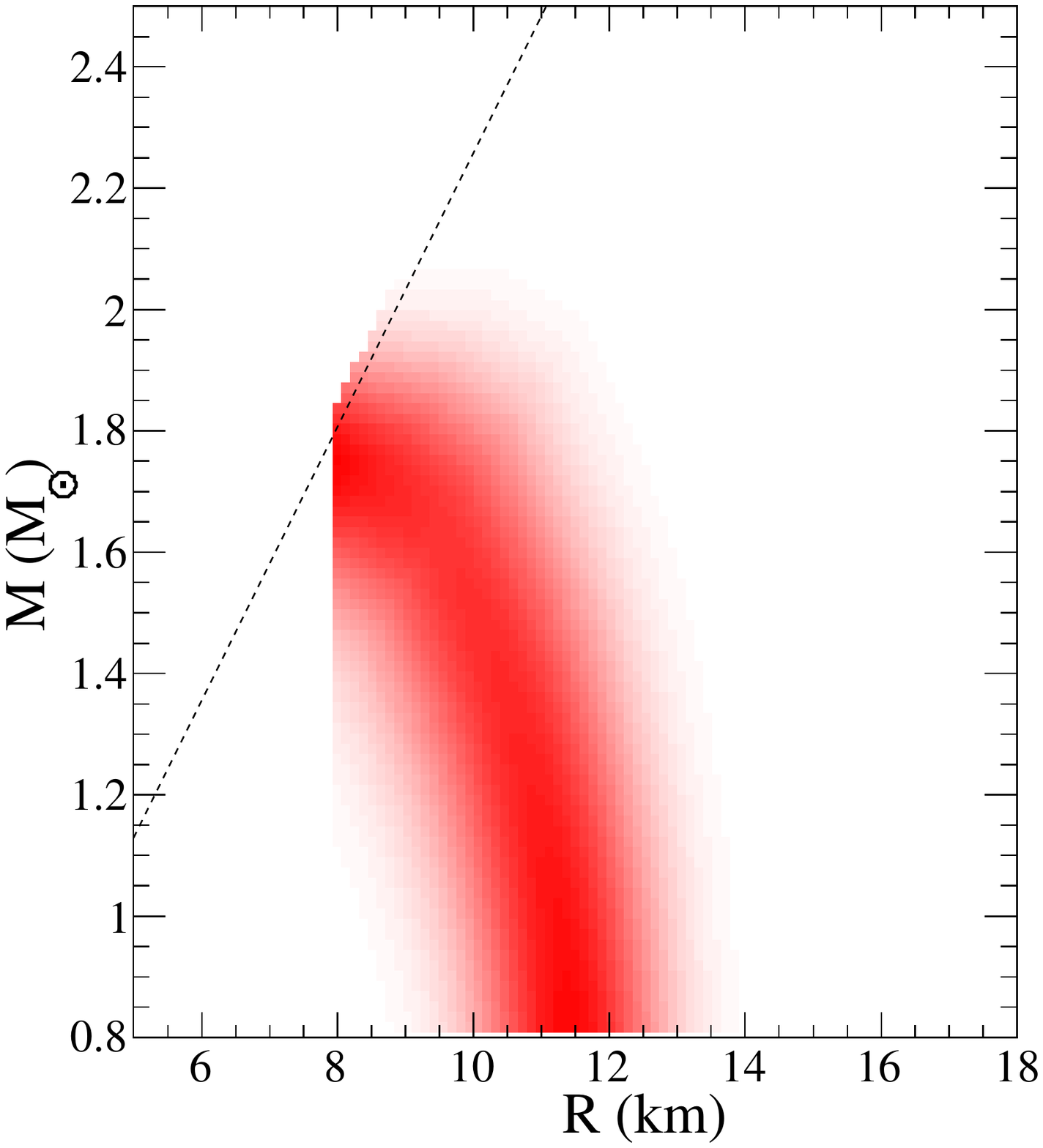}}
\put(0,-220){\includegraphics[width=.45\columnwidth]{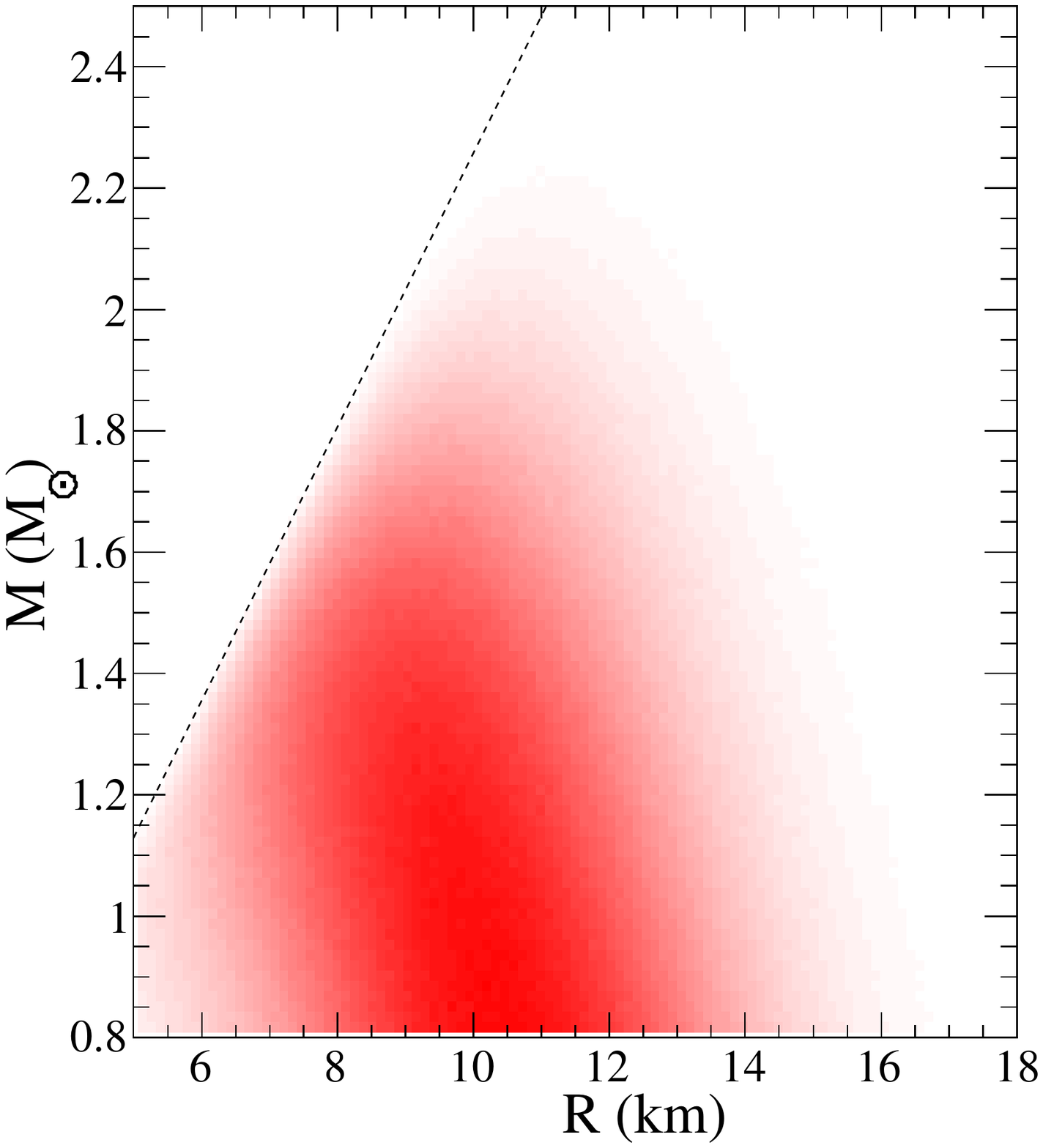}}
\put(140,-220){\includegraphics[width=.45\columnwidth]{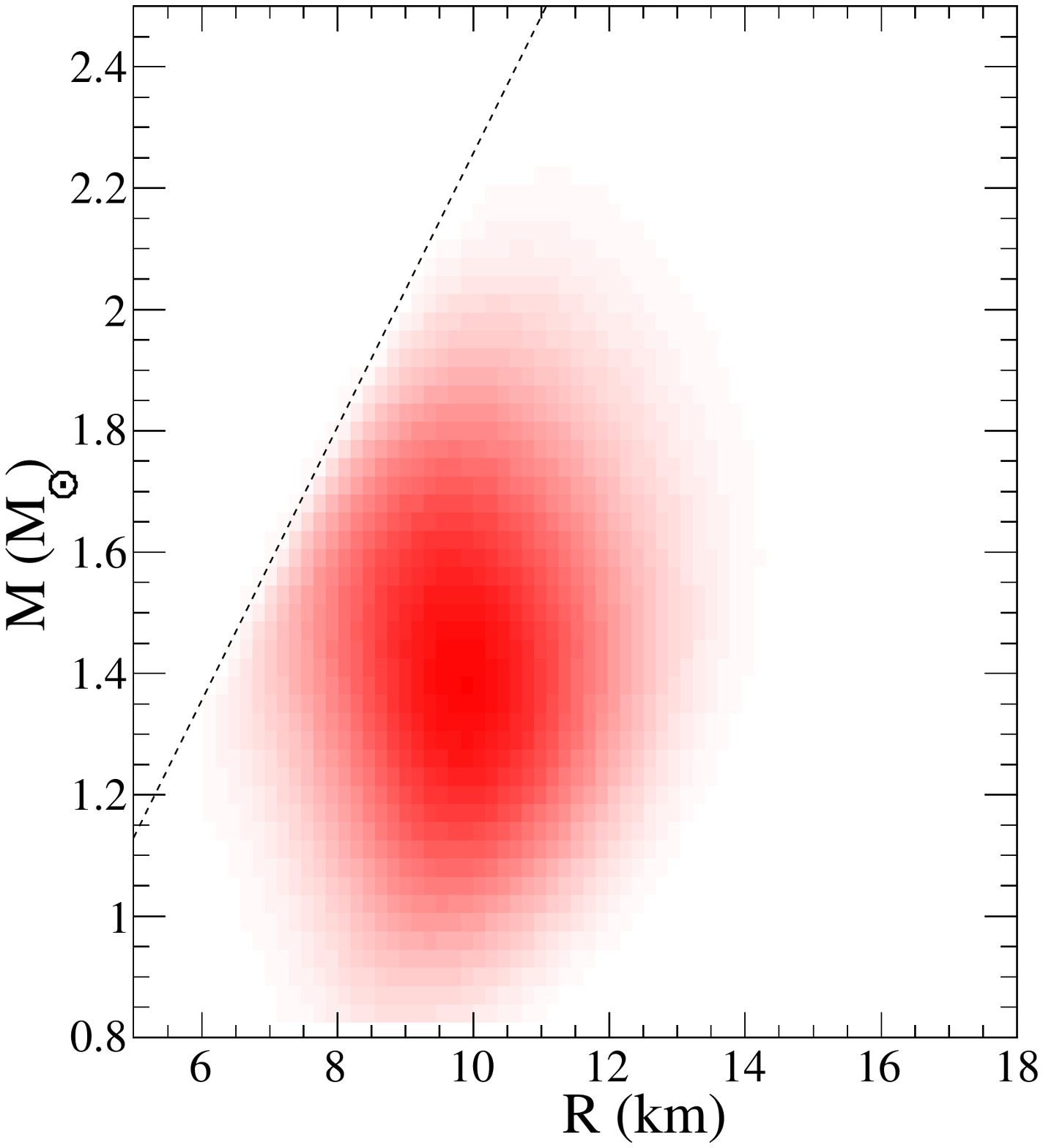}}
\put(280,-220){\includegraphics[width=.45\columnwidth]{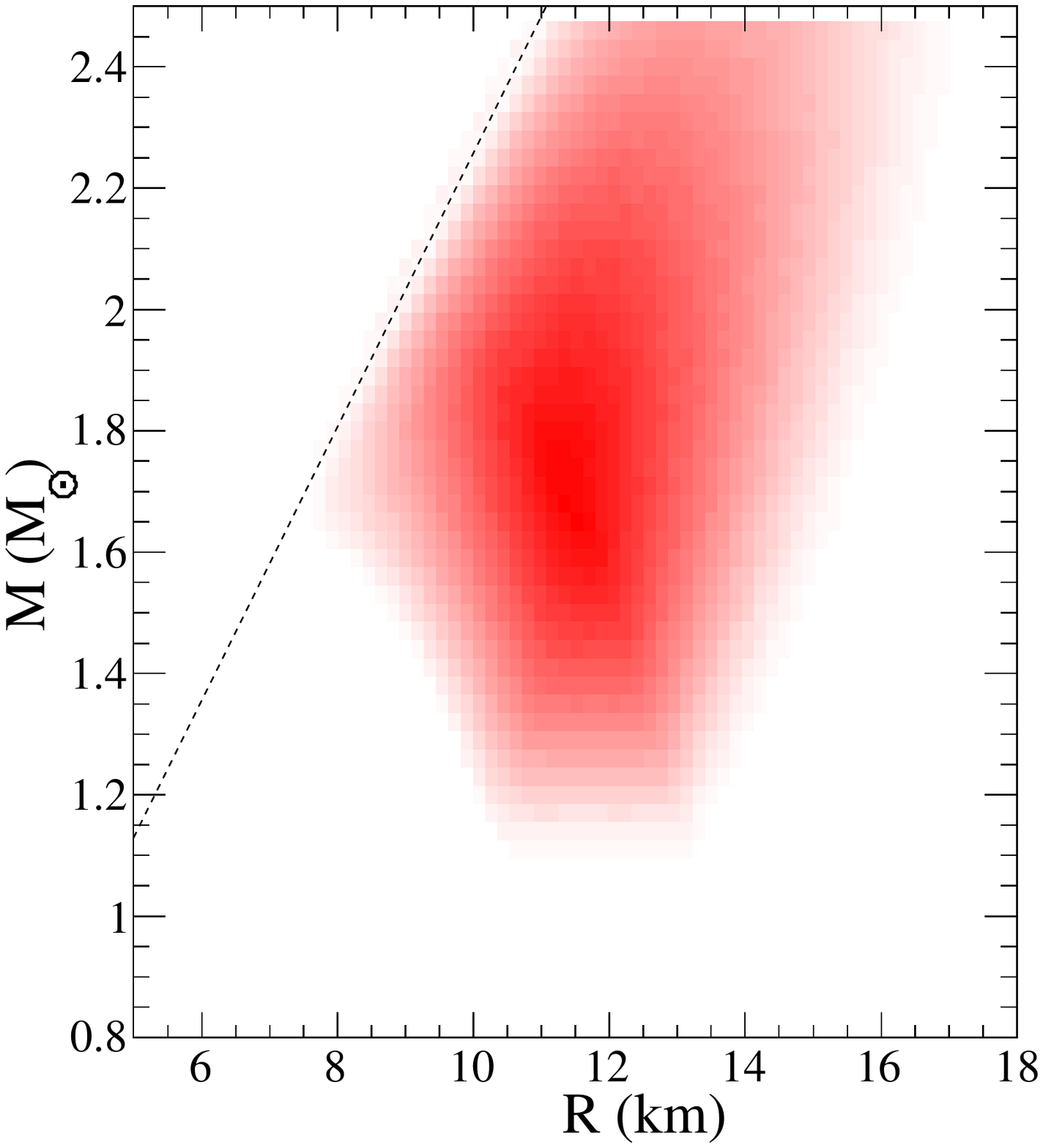}}
\put(0,-380){\includegraphics[width=.45\columnwidth]{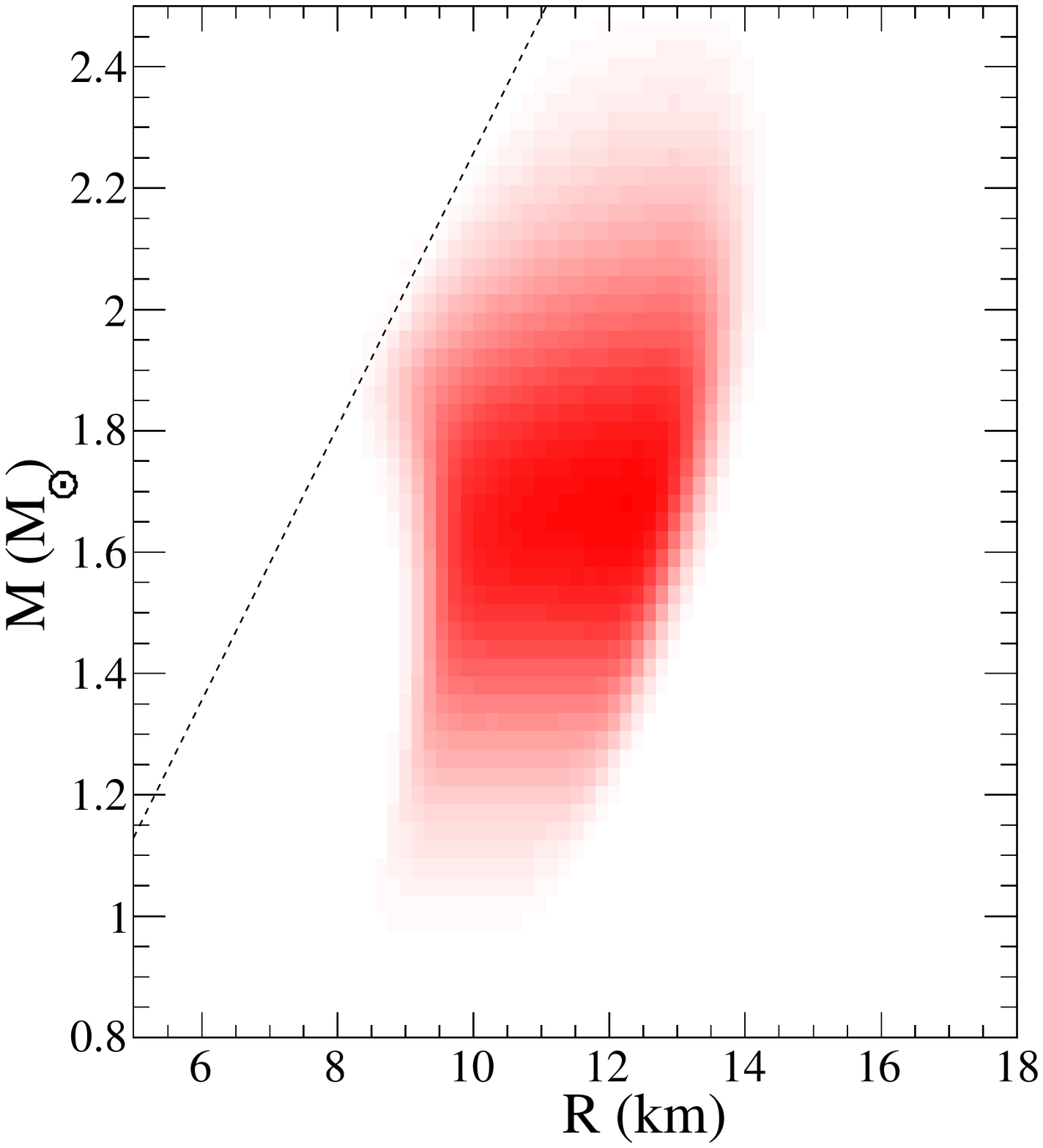}}
\put(140,-380){\includegraphics[width=.45\columnwidth]{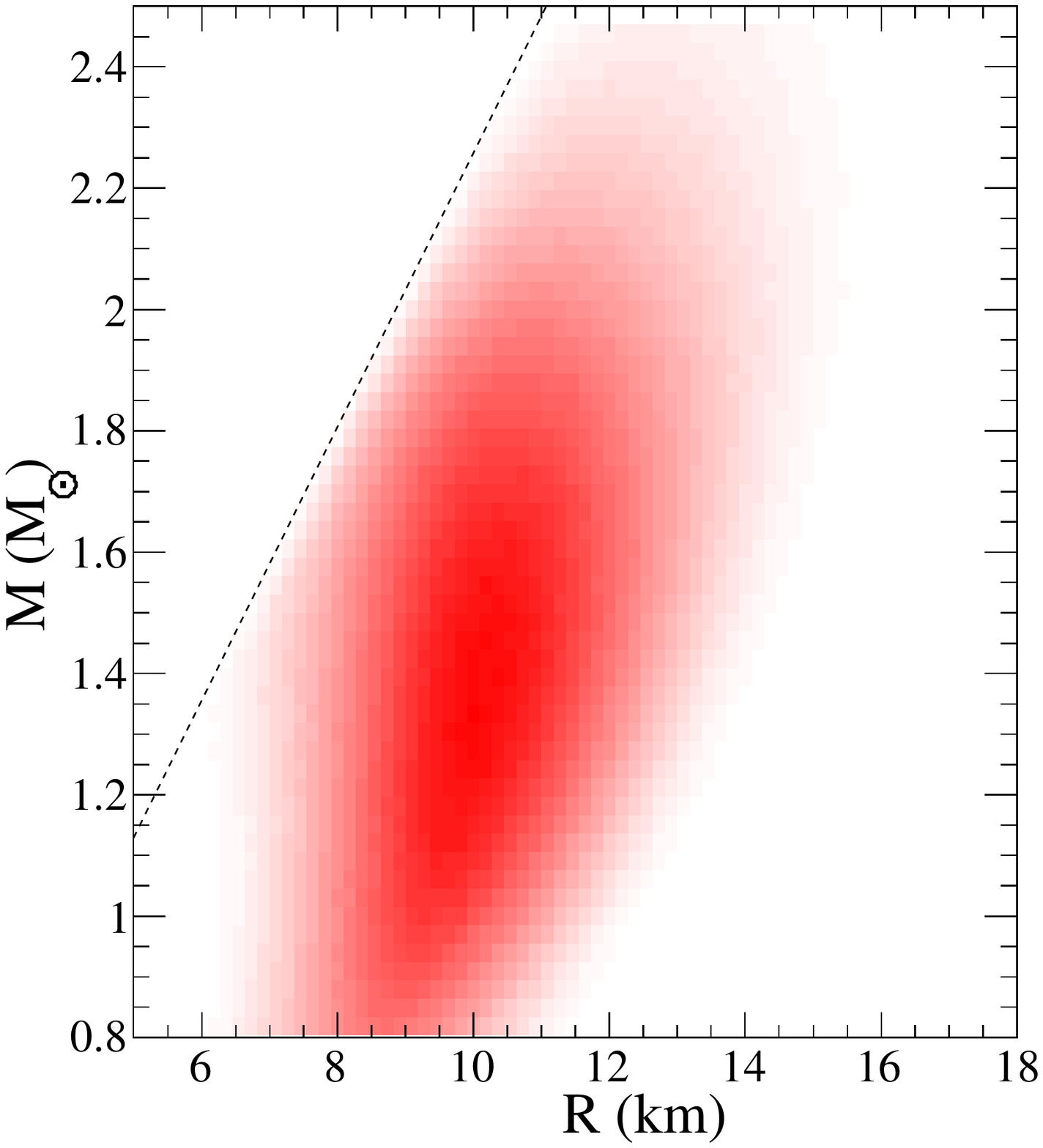}}
\put(20,60){ M13}
\put(170,60){X7}
\put(305,60){$\omega$ Cen}
\put(30,-100){U24}
\put(170,-100){1745}
\put(305,-100){1608}
\put(30,-260){1820}
\put(170,-260){1731}
\end{picture}
\vspace*{12.75cm}\caption{$M-R$ probability densities of neutron stars from (a) four quiescent
  low-mass X-ray binaries in globular clusters and (b) four
  photospheric radius expansion burst sources (incorporating the
  possibility that $R_{ph}\ge R$).  The diagonal lines represent
  causality limits.  Figure reproduced courtesy of A. W. Steiner.}
\label{therm}
\end{figure}

The best-studied isolated neutron star is RX J1856-3754 \cite{W96},
which is the only one close enough to have an accurate parallactic
distance: $D=120\pm8$ pc \cite{W10}.  Fitting both Rosat X-ray and
optical spectra with non-magnetic heavy-element atmospheres \cite{P02}
gives $R_\infty=16.1\pm1.8\mathrm{~km}$ and redshift $z=0.37\pm0.03$
for this distance.  These values lead to $M=1.86\pm0.23$ M$_\odot$ and
$R=11.7\pm1.3$ km.  A later analysis with Chandra data
found $z\simeq0.3$ and $R_\infty\simeq15.8$ km \cite{Walter04}.
However, these models predict spectral features that are not observed, and other observations indicate a
substantial magnetic field of order $5\times10^{12}$ G.  Burwitz et al. \cite{Burwitz03} and
Ho et al. \cite{H07} proposed models with a highly-magnetized,
condensed surface, yielding $R_\infty>13$ km in the first case and
$R_\infty\simeq14.6\pm1.0$ km and $z\simeq0.22$ in the second, for
$D=120$ pc.  Such models require a trace H
atmosphere with a finely- tuned mass whose origin is unclear.  Ho et al.'s values imply
$M=1.33\pm0.09$ M$_\odot$ and $R=11.9\pm0.8$ km (errors include only the
distance uncertainty).  Although the predicted masses of these two approaches differ, the predicted radii are nearly identical.

Many neutron stars are transients in which accretion 
proceeds intermittently; the accretion episodes are separated by long
periods of quiescence.  During accretion, compression of
matter in the crust induces nuclear reactions \cite{Haensel90} that
release heat.  When the accretion ceases, the heated crust cools and
radiates an observable thermal
spectrum \cite{Brown98}. Accretion is also believed to suppress the
surface magnetic field, which seems to be confirmed by the lack of evidence, such as pulsations or cyclotron spectral features, for a
magnetic field significant enough to distort the star's spectrum.  Because the timescale for
heavier nuclei to sink below the photosphere is short, the spectra can be fitted
with well-understood unmagnetized H atmosphere models.  Modeling 
can be used to reliably infer the apparent angular emitting area, and,
possibly, the surface gravity \cite{Heinke06}, both of which are functions of $R$ and $M$.  Figure \ref{therm} summarizes the probability distributions of $M$ and $R$ for four sources in
globular clusters \cite{Heinke06,Webb07}.

\subsection{Photospheric Radius Expansion Bursts}

Type I X-ray bursts are the result of thermally unstable helium (or, in
some cases, H) ignition in the accreted envelope of a neutron
star \cite{SB04}. The ignition generates a
thermonuclear explosion that is observed as an X-ray burst with a
rapid rise time ($\sim 1\,\mathrm{s}$) followed by cooling decay
($\sim 10\textrm{--}100\,\mathrm{s}$).  If the burst is sufficiently
luminous, radiation pressure drives surface layers and the photosphere outwards to larger
radii, and the flux at the photosphere approaches (to within a few
percent) the Eddington value,
\begin{equation}
F_{Edd}={cGM\over\kappa D^2}\sqrt{1-2\beta_{ph}}
\label{eq:Fedd}
\end{equation} 
for which radiation pressure balances gravity; the flux then decreases. The opacity of the lifted material is $\kappa$ and $\beta_{ph}=GM/(R_{ph}c^2)$,  where $R_{ph}$ is the radius of the photosphere.  The blackbody
temperature $T_{bb,\infty}$ also reaches a maximum, marking touchdown as the lifted material falls to the surface. At that time, the observed angular
area of the photosphere is
\begin{equation}
A={F_{\infty}\over T_{bb,\infty}^4}=f_c^{-4}\left({R_{bb,\infty}\over D}\right)^2,
\label{eq:a}
\end{equation}
where $f_c$ is the color-correction factor approximating the effects of the atmosphere in distorting the observed temperature from the effective blackbody temperature.  After touchdown, $R_{bb,\infty}=R_{ph}/\sqrt{1-2\beta_{ph}}$
stabilizes as the observed flux and temperature decrease.   The standard view is that $R_{ph}=R$; that is, the photosphere lies very close to the surface.

The observed parameters are $F_{Edd}, A$, and $D$.  Atmosphere models for different compositions and effective temperatures provide $\kappa$ and $f_c$. 
We define
\begin{equation}
\alpha \equiv \frac{F_{Edd}}{\sqrt{A}}
\frac{\kappa D}{c^3f_c^2}=\beta\sqrt{1-2\beta}\sqrt{1-2\beta_{ph}}, 
\label{eq:agr1}
\end{equation}
and
\begin{equation}
\gamma \equiv \frac{Ac^3f_c^4}{F_{Edd}\kappa}= 
\frac{R}{\beta(1-2\beta)\sqrt{1-2\beta_{ph}}}. 
\label{eq:agr}
\end{equation}
Note that $\alpha\gamma=R_\infty$.  $M$ and $R$ can therefore be determined if
$\alpha$ and $\gamma$ are measured, assuming knowledge of $D, \kappa$, and $f_c$.  If $R_{ph}=R$, then
$\alpha=\beta(1-2\beta)$ which has two real-valued solutions if
$\alpha\le1/8$ and none otherwise.  The range of $\gamma$ values for the observed sources is not large.

\"Ozel and collaborators \cite{Ozel09,Ozel10} have studied the
bursters EXO 1745-248, 4U 1608-522, 4U 1820-30 and KS 1731 and have estimated $F_{Edd}$ and $A$ for each.  Coupled with estimates for $D$, these authors found
$\alpha\simeq0.13\pm0.02, 0.18\pm0.06$, and $0.17\pm0.02$, respectively. 
Thus, assuming that $R_{ph}=R$, no real-valued solutions exist for the
centroids of the data.  However, real solutions can be found in the
tails of the error distributions for $F_{Edd}$, $A$, and $D$, if one
also takes into account theoretical uncertainties in $f_c$ and
$\kappa$ which are not negligible.  Monte Carlo sampling of
$F_{Edd},A,D,f_c$, and $\kappa$ within their estimated errors,
accepting only trials with real solutions, allows the generation of
probability distributions for $M$ and $R$.  Because requiring real
solutions severely restricts the allowed values of input variables,
the probability distributions in $M$ and $R$ have smaller errors (of
order 5\%) than the observational inputs. Another consequence is that
the mean values of $M$ and $R$ do not greatly vary from source to
source.  Relatively small values for the radius ($R\simeq8-10$ km) are
found.

The Monte Carlo sampling has extremely small fractional
acceptance rates: 0.13, 0.0002, and $2\cdot10^{-8}$ for the first three
of the above sources, for example.  This finding casts doubt on the model's
validity.  Accepted trials are forced into a small region where
$\alpha\simeq1/8$, which, coupled with the fact that $\gamma$ has an
intrinsically small variation among sources, explains the small errors and
overlap of $M$ and $R$ values for different sources.

It is likely this model has systematic uncertainties that have not been
considered.  One possibility, suggested by Steiner et al.~\cite{SLB10}, is that
the photospheric radius at the point where the Eddington flux is
measured might not concide with the stellar surface.  In the extreme
case in which $R_{ph}>>R$, one has $\alpha=\beta\sqrt{1-2\beta}$, which has
one physical real-valued solution if $\alpha\le3^{-3/2}\simeq0.192$
and none otherwise.  If one allows for the possibility that $R_{ph}\ge R$,
Monte Carlo acceptance rates increase dramatically \cite{SLB10}.
Predicted errors for $M$ and $R$ also increase, which leads to 1-$\sigma$
errors of order 15\% (Figure \ref{therm}).  Another
consequence is that estimated radii are, in each case,
approximately 2 km larger than \"Ozel et al. obtained.

It has been proposed \cite{SPRW11}, conversely, that the short X-ray
bursts (used in References \cite{Ozel09} and \cite{SLB10}) have color-correction factors
that change significantly during the burst, which could possibly explain
inconsistencies such as the lack of real solutions.  By using only long
bursts and different model atmospheres, Suleimanov et
al.~\cite{SPRW11} found distinctly larger radii: $R>14$ km.  However,
this result is incompatible with studies of nuclei and neutron matter,
as discussed in Section \ref{sec:lab}.  This result is also inconsistent with studies of
sub-Eddington X-ray bursts from GS 1826-24 \cite{Zamfir12}.
Further modeling of X-ray bursts is
obviously warranted.

\section{FROM OBSERVATIONS TO THE EQUATION OF STATE\label{sec:m-r}}

With several $M-R$ probability distributions, one can investigate
constraints on the overall $M-R$ relation, and the EOS
($p-\varepsilon$ relation) can be inferred by inversion of
Equation~(\ref{TOV}).  Steiner et al. \cite{SLB10} assumed a
parameterized form for the EOS and determined the most
likely parameter values from a Bayesian analysis of the $M-R$
probability distributions by using Markov chain Monte Carlo integration
techniques.  The possibility that $R_{ph}\ge R$ for the X-ray burst sources was
incorporated.  Below the transition density between the core and crust
of a neutron star, $n_{\mathrm{trans}}\simeq n_s/4$, the EOS was
assumed to be known because it is dominated by electronic and lattice
pressures.  In the vicinity of $n_s$, a standard expansion of the
nucleon energy 
in powers of $n-n_s$ and $1-2x$ was assumed, as in Equation~\ref{eose}.
The symmetry energy is parameterized as
\begin{equation}\label{eq:ldeos}
S_2(u)=S_ku^{2/3}+S_pu^\gamma.
\end{equation}
The kinetic part of the
symmetry energy, $S_k\simeq13$ MeV, was held fixed, but the
compressibility and skewness coefficients, $K_o$ and $K^{\prime}_o$, and the
terms describing the potential part of the symmetry energy, $S_p$ and 
$\gamma$, were taken as parameters.  

Above the density $\varepsilon_1$, two polytropic relations with
exponents $\gamma_1$ and $\gamma_2$, separated at the density
$\varepsilon_2$, were assumed.  Although  
nuclear masses can establish strong correlations among some parameters,
such as between $S_v$ and $\gamma$, these were ignored.  The ranges of
parameters were constrained to satisfy causality and $M_{max}>1.66$ M$_\odot$, although raising this limit to $M_{max}>1.93$ M$_\odot$ in order to comply with observations of PSR J1614-2230 only slightly affected the results \cite{steiner12}.

\begin{figure}[h]
\vspace*{4cm}
\begin{picture}(6,12)
\put(-20,0){\includegraphics[width=6.75cm,trim=0 0.cm 5.2cm 0.cm,clip]{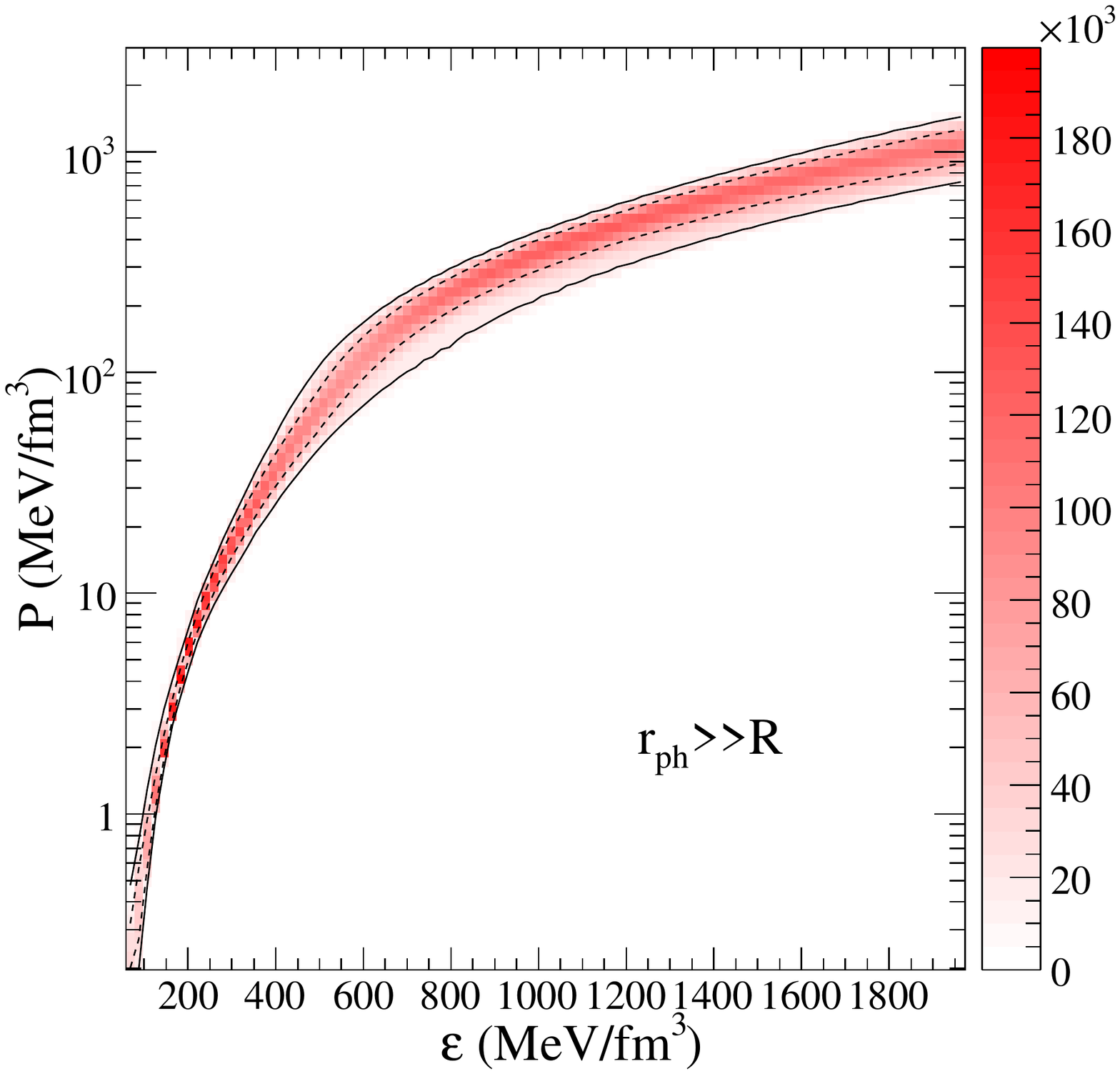}}
\put(150,-125){\includegraphics[width=9.75cm]{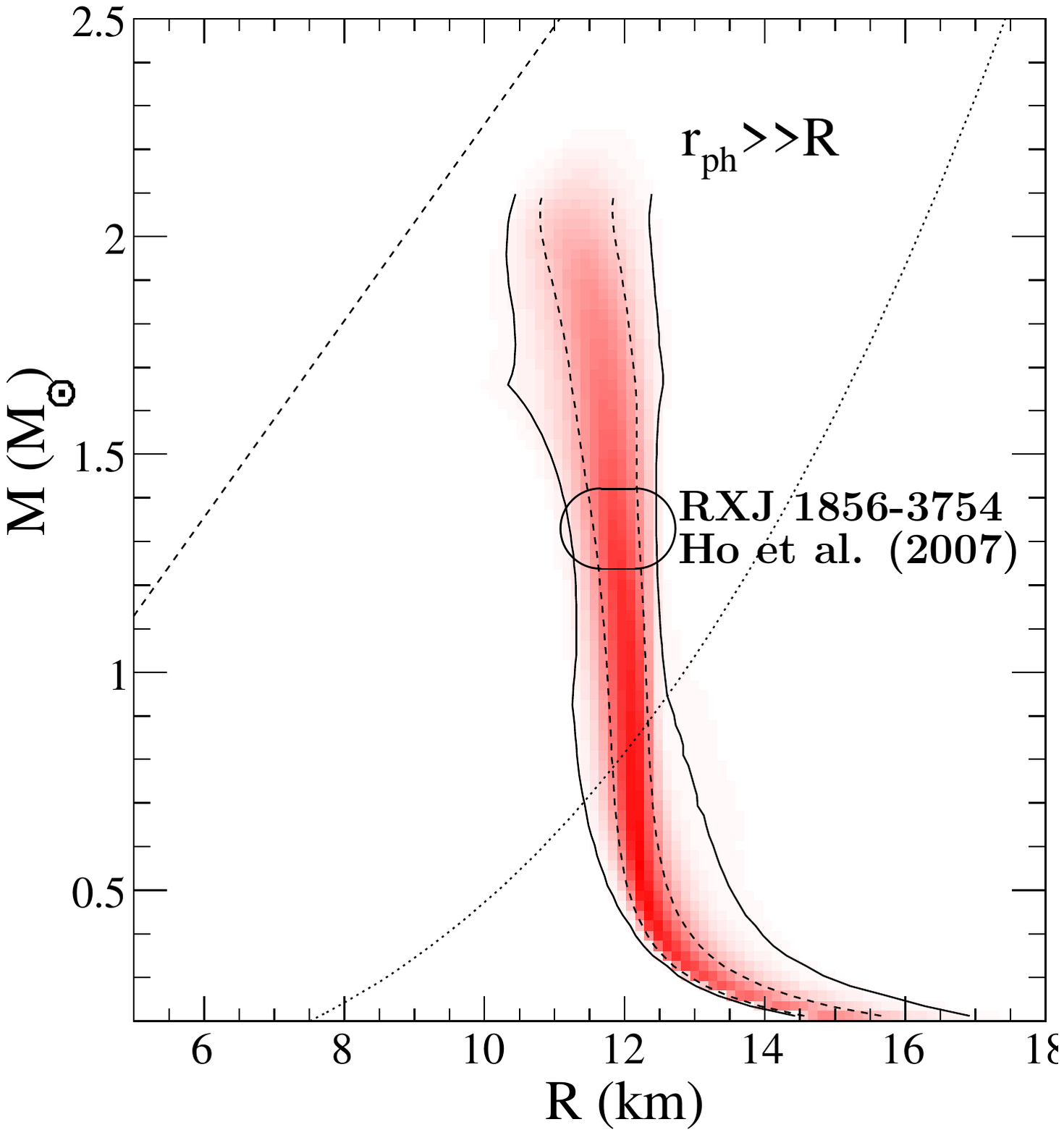}}
\end{picture}
\caption{(a) Probability distributions for pressure as a
function of energy density using the $M-R$ probability distributions from
Figure \ref{therm}.  (b) Probabilty distributions for the $M-R$ curve.  The diagonal dashed line is the causility limit,
and the curved dotted line is the 716 Hz rotation constraint.  1-$\sigma$ and 2-$\sigma$ contours are shown as dashed and solid lines, respectively.
Also shown is the estimated mass and radius error region, including only distance errors, for RXJ1856-3754 \cite{H07}.  Figures adapted from Reference~\cite{SLB10}.}
\label{pemr}
\end{figure}
The Bayesian analysis determines the most likely $M-R$ relations
(Figure~\ref{pemr}b) and the most likely values for the EOS
parameters.  Figure \ref{pemr}a displays the most likely ranges for
the EOS.  1.4-M$_\odot$ neutron stars are most likely to have radii of
order $11.5\pm1$ km.  These small radii are a consequence of a rather
soft nuclear symmetry energies, $\gamma=0.3\pm0.1$.  Note that the
results \cite{P02,H07} from spectral modeling of RX J1856-3754 are
also consistent with the derived $M-R$ relation (Figure~\ref{pemr}).
Although the EOS is relatively soft in the vicinity of $\rho_s$, the
high-density EOS must stiffen: The Bayesian analysis indicates that
the neutron star maximum mass is $M_{max}=2.05\pm0.11$ M$_\odot$.

Interestingly, the predicted most-likely values of three
nuclear matter parameters ($190{\rm~MeV}<K<260{\rm~MeV}$,
$26{\rm~MeV}<S_v<34{\rm~MeV}$, and $0.2<\gamma<0.4$) are compatible
with experimental information.  (The skewness $K^\prime$ is neither
experimentally nor observationally well-constrained.)  $K$ is
well-constrained by monopole resonances to be
$230{\rm~MeV}<K<250{\rm~MeV}$ (see Reference~\cite{Piekarewicz10} for a
review).  The symmetry parameters are also consistent with
experimental information (see Section \ref{sec:lab}).  There appears to be a
remarkable convergence between experimental and astrophysical
estimates for the neutron star EOS.

These estimates were made on the basis of relatively poor radius-mass
information from only seven or eight neutron stars.  The results are somewhat
sensitive to how photospheric radius expansion bursts are modeled.  On
the one hand, if one assumes $R_{ph}=R$, the predicted radii would
become $\approx2$ km smaller \cite{Ozel10}; however, in this case, the
$M_{max}\ge1.93$ M$_\odot$ constraint from PSR J1614-2230 can no
longer be satisfied \cite{SLB10}.  On the other hand, if the burst
data are excluded, the inferred $M-R$ relation is essentially
unaltered from the baseline results achieved incorporating the burst sources
with the possibility that $R_{ph}\ge R$ (A. W. Steiner, private
communication).  In this case, the observed $M-R$ results are largely
a consequence of the wide range of observed $R_\infty$ values of the
quiescent globular cluster sources (Figure \ref{therm}) which forces
the $M-R$ curve to enter its vertical trajectory at relatively
small radii ($\sim11$ km).  Otherwise, the existence of both small and
large observed values of $R_\infty$ would not be compatible with
realistic masses.

$M-R$ information has also been inferred from pulse-shape modeling of
X-ray bursts.  Although predictions from observations of individual
sources have large errors, Leahy et al. \cite{Leahy11} concluded that
only an $M-R$ curve with a constant radius of $R\sim12$ km for 1
M$_\odot<M<2.3$ M$_\odot$ would be consistent with observations of all
sources studied, namely XTE J1807-294, SAX J1808-3658, and XTE
J1814-334.  This result is remarkably similar to the conclusions drawn
in Reference \cite{SLB10} (Figure~\ref{pemr}).

Nevertheless, Suleimanov et al's \cite{SPRW11} study of longer X-ray
bursts implied considerably larger radii: $R\simge14$ km.  Those
results are further supported by spin-phase resolved spectroscopy of
isolated neutron stars \cite{Hambaryan11} that yield small neutron
star redshifts: $z\simeq0.16$.  This value for $z$, coupled with
realistic neutron star masses ($M>1.2$ M$_\odot$), implies that $R>14$
km.  Because these results are also incompatible with available
experimental information discussed in the next section, it is
important to resolve these differences.  More
sophisticated modeling of photospheric radius expansion bursts and
neutron star atmospheres, together with refinements of distances, will
be required for additional progress.

\section{LABORATORY CONSTRAINTS\label{sec:lab}}
The most significant aspects of neutron star structure impacted by the
EOS are the maximum mass and the typical radius of intermediate stars.
The former is mostly sensitive to the EOS beyond three times the
saturation density.  Figure~\ref{mr} shows that this density is the
minimum central density of stars greater than approximately 1.5
M$_\odot$, whereas the central densities at the maximum mass can
approach $7-8~n_s$.  However, the radii of stars with mass $\simle1.5$
M$_\odot$ are completely determined by the EOS below $3n_s$, and, in
particular, by $dS/du$ in the vicinity of $n_s$.
  
\subsection{Nuclear Symmetry Energy}
The nuclear symmetry energy $S_2(u)$ is often described by the liquid drop parameters
\begin{eqnarray}
S_v&=&S_2(1),\nonumber\\
L&=&3u\left({\partial^3 e\over\partial u\partial^2x}\right)_{u=1,x=1/2}=3u\left({dS_2\over du}\right)_{u=1},\\
K_{sym}&=&9u^2\left({\partial^4 e\over\partial u^2\partial^2x}\right)_{u=1,x=1/2}=9u^2\left({d^2S_2\over du^2}\right)_{u=1}\nonumber
\label{eospar}\end{eqnarray}
For example, in the parameterization Equation~(\ref{eq:ldeos}),
$L=2S_k+3\gamma(S_v-S_k)$ and
$K_{sym}=-2S_k+9\gamma(\gamma-1)(S_v-S_k)$.  There are limited
constraints on the total incompressibility of nuclei from measurements
of the giant isoscalar monopole resonance \cite{Li10}, isotopic
transport ratios in medium-energy heavy-ion collisons \cite{Chen09},
and neutron skin data from anti-protonic atoms \cite{Centelles09}.
However, extraction of the symmetry contribution to the total
incompressibility of nuclei and its further separation into bulk
($K_{sym}$) and surface portions is still problematic.


In contrast, there are several methods of experimentally determining
$S_v$ and $L$, including (a) nuclear mass fits, (b) neutron skin
thicknesses, (c) dipole polarizabilities, (d) dynamics in heavy-ion
collisions, (e) giant and pygmy dipole resonances, and (f) isobaric
analog states.

The most accurate data are obtained from fitting nuclear masses.
However, optimizing the symmetry parameters of nuclear models to
experimental masses cannot uniquely constrain them.  Rather, the
parameters are highly correlated
\cite{Lattimer92,oyamatsu03,danielewicz03,SPLE05,Kortelainen10}.

It is straightforward to demonstrate the existence of this correlation
using the liquid drop model \cite{weizsacker35}.  Although the
specific results are sensitive to the inclusion of shell and pairing
terms, Coulomb diffusion and exchange terms, a Wigner term, and the
neutron skin, these effects are inconsequential for this
demonstration.  Neglecting these contributions, the symmetry energy of
a nucleus to lowest order is $E_{mod,A}=I^2(S_vA-S_sA^{2/3})$, where
$I=(N-Z)/(N+Z)$, and consists of volume and surface terms.
The volume parameter $S_v$ is the same as the bulk parameter in
Equation~\ref{eospar}.  $S_s$ is the surface symmetry energy parameter
which is sensitive to the density dependences of both $e$ and $S$ and
thus depends on $S_v$ and $L$ (see below).
Minimizing the difference between the model and experimental symmetry
energies for nuclei with measured masses---that is, $\chi^2=\sum_i(E_{exp,i}-E_{mod,i})^2/({\cal N}\sigma^2)$, where
${\cal N}$ is the total number of nuclei and $\sigma$ is a nominal error---establishes a correlation between $S_v$ and $S_s$.
This correlation can be visualized as 
a 1-$\sigma$ confidence ellipse: a $\chi^2$ contour that is one unit
larger than the minimum value of $\chi^2$.  The shape and orientation of this ellipse are
determined by the second derivatives of $\chi^2$ at the
minimum,
\begin{equation}\label{second}
[\chi_{vv},~\chi_{vs},~\chi_{ss}]\sigma^2={2\over{\cal N}}\sum_iI_i^4[A_i^2,~-A_i^{5/3},~A_i^{4/3}]\simeq[61.6,~-10.7,~1.87],
\end{equation}
where $\chi_{vs}=\partial^2\chi^2/\partial S_v\partial S_s$, and so on.  In
the simple liquid drop case, these derivatives do not depend on the location of the
minimum.  The confidence ellipse has an orientation
$\alpha=(1/2)\tan^{-1}|2\chi_{vs}/(\chi_{vv}-\chi_{ss})|\simeq9.8^\circ$
with respect to the $S_s$ axis; the ellipse's semimajor axes are the error widths are
$\sigma_v=\sqrt{(\chi^{-1})_{vv}}\simeq2.3$ MeV and $\sigma_s=\sqrt{(\chi^{-1})_{ss}}\simeq13.2$ MeV,
where $(\chi^{-1})$ is the matrix inverse and we used $\sigma=1$ MeV.

In order to convert the correlation between $S_v$ and $S_s$ into a
correlation between $L$ and $S_v$, one needs to estimate how $S_s$
depends on $S_v$ and $L$.  Treating the nuclear surface as a
plane-parallel interface, $S_s$ can be expressed simply as the ratio
of two integrals over the nuclear density $u=n/n_s$ of a symmetric
system \cite{krivine83,danielewicz03,SPLE05}:
\begin{equation}
S_s={E_sS_v\over2}{\int_0^{1}u^{1/2}[S_v/S_2(u)-1][e(u,1/2)+B]^{-1/2}du\over\int_0^1u^{1/2}[e(u,1/2)+B]^{1/2}du}.
\end{equation}
Here $E_s\simeq19$ MeV is the liquid drop surface energy parameter (also found
from fitting nuclear masses).  The solution depend on the functional forms of $S$ and $e$ and is generally
nontrivial, but an analytical result is found if we approximate
$S_2(u)\simeq S_v+L(u-1)/3$ and $e(u,1/2)+B\simeq K(u-1)^2/18$:
\begin{equation}
{S_s\over S_v}\simeq{135E_s\over2K}\left[1-\left[{3S_v\over L}-1\right]^{1/2}\tan^{-1}\left(\left[{3S_v\over L}-1\right]^{-1/2}\right)\right].
\end{equation}
For example, for $L/(3S_v)\simeq1/3,~1/2$, or $2/3$, one finds $S_s/S_v\simeq0.7,~1.2$, or 1.8, which shows the basic trend.
For various interactions constrained to broadly reproduce nuclear masses in Thomas-Fermi calculations, an approximate relation can be found \cite{Lim12}:
\begin{equation}\label{ssl}
{S_s\over S_v}\simeq0.6461+{S_v\over97.85{\rm MeV}}+0.4364{L\over S_v}+0.0873\left({L\over S_v}\right)^2.
\end{equation} 
By transforming to $S_v-L$ space, one finds the confidence ellipse in $S_v-L$ space has $\sigma_v\simeq2.3$ MeV and $\sigma_L\simeq20$ MeV.

In practice, the liquid droplet model \cite{MS69}, which accounts for
varying neutron/proton ratios within the nucleus, provides an improved
fit to nuclear masses.  Its symmetry energy is
$E_{mod,A}=I^2_iS_vA(1+S_sA^{-1/3}/S_v)^{-1}$ and the above
methodology can be used to determine the confidence ellipse.  One
finds $\sigma_v\simeq3.1$ MeV and $\sigma_L\simeq47$ MeV (these values now depend
on the location of the minimum, $S_{v0}\simeq30.5$ MeV and $L_0\simeq62$ MeV).  

Modern microscopic models of nuclei are a vast improvement over liquid
drop and droplet models.  Nevertheless, they predict an $S_v-L$
correlation in substantial agreement with the liquid droplet model or
Thomas-Fermi calculations \cite{oyamatsu03,Lim12}.  The correlation
was recently studied by Kortelainen et al. \cite{Kortelainen10} using
the Universal Nuclear Energy Density Functional \cite{Bertsch07}.
Fitting nuclear masses and charge radii resulted in a confidence
ellipse with the properties $S_{v0}\simeq30.5$ MeV, $L_0\simeq45$ MeV,
$\sigma_v=3.1$ MeV, and $\sigma_L=40$ MeV, and an orientation
indistinguishabe from that predicted by the liquid droplet model and
Thomas-Fermi calculations.  The correlation of these parameters was
over 97\% which is illustrated in Figure~\ref{correl} for the 90\%
confidence contour.  We additionally note that the latest microscopic
finite-range droplet model \cite{Moller12}, when optimized to nuclear
masses, predicted best-fit values of $S_{v0}=32.5$ MeV and $L_0=70$
MeV.  This result lies precisely on the Kortelainen et al. correlation
line within their confidence ellipse (Figure~\ref{correl}).  This
correlation is therefore robust and model- independent.

\begin{figure}[h]
\includegraphics[angle=270,width=14cm,angle=90]{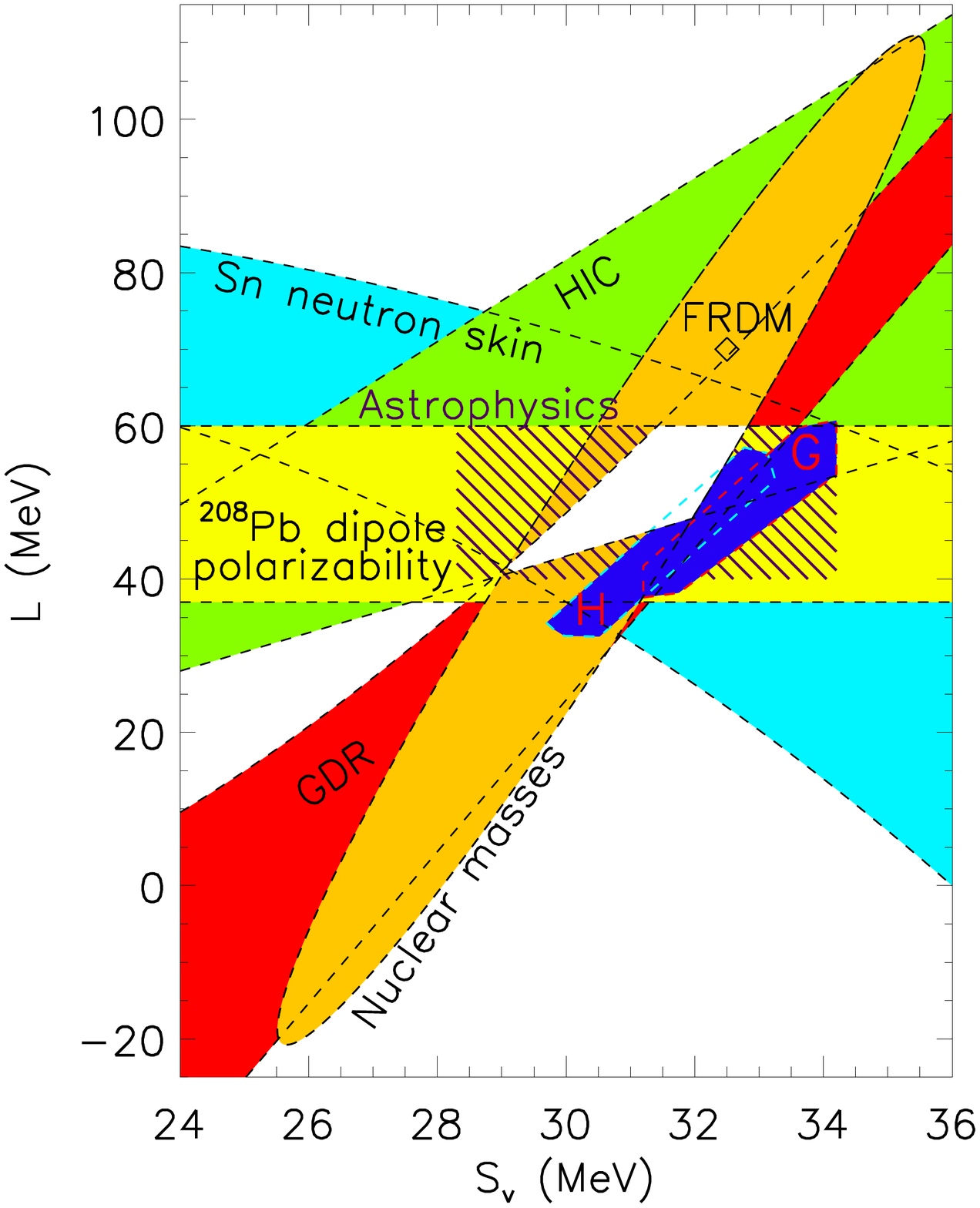}
\caption{Summary of constraints on symmetry energy parameters.  The
  filled ellipsoid indicates joint $S_v-L$ constraints from nuclear
  masses \cite{Kortelainen10}.  The finite-range droplet model (FRDM) fit \cite{Moller12} is indicated with a diamond.  The filled bands show constraints from
  neutron skin thickness of tin (Sn) isotopes \cite{Chen10}, isotope
  diffusion in heavy-ion collisions (HIC)\cite{Tsang09}, the dipole
  polarizability of $^{208}$Pb \cite{Piekarewicz12}, and giant dipole
  resonances (GDR) \cite{Trippa08}.  The hatched rectangle shows
  constraints from astrophysical modeling of $M-R$ observations
  \cite{SLB10}.  The two filled regions show neutron matter constraints
  (H is from Reference \cite{HLPS10} and G is from Reference
  \cite{GCR12}).  The white area is the experimentally-allowed overlap
  region.  Figure adapted from Reference \cite{Lim12}.}
\label{correl}
\end{figure}

This correlation has been used to constrain nuclear parameters in
supernova simulations which explored the effects of variations in the
nuclear symmetry energy (see, for example, Reference \cite{swesty94}).
Coupled with other nuclear observables, this correlation can become even more
effective.  One possibility concerns the neutron skin thickness of
neutron-rich nuclei.  Neglecting Coulomb effects, the difference
between the mean neutron and proton surfaces predicted by the liquid droplet
model \cite{MS69}
\begin{equation}\label{skin}
t_{np}={2r_o\over3}{S_sI\over S_v+S_sA^{-1/3}}.
\end{equation}
Therefore, neglecting Coulomb effects, the neutron skin thickness of
any particular isotope is predicted to be primarily a function of
$S_s/S_v$ and hence, through Equation~(\ref{ssl}), $L$ and $S_v$.  Values of
$t_{np}$ have been measured, typically with 30-50\% errors, for approximately
two dozen isotopes.  A recent study \cite{Chen10} fitting the neutron
skin thicknesses of tin isotopes, in which differential isotopic measurements
reduce errors, found a correlation between $S_v$ and $L$ that is
nearly orthogonal to the mass-fit correlation (Figure~\ref{correl}).
Additionally, neutron skin measurements of anti-protonic atoms results
in a range of 25 MeV to 70 MeV \cite{Centelles09} for $L$ (this is
consistent with the tin study, but not displayed in Figure~\ref{correl}
for clarity).  The first results \cite{Abrahamyan12} from the PREX
experiment to measure the neutron radius of $^{208}$Pb have indicated
a range of 35 MeV $<L<$ 262 MeV.  The mean value is discrepant with
other results for neutron skins, but the errors are still very large.

Figure~\ref{correl} shows an additional constraint obtained from
isospin diffusion in heavy-ion collisions \cite{Tsang09}.  

Tamii et al. \cite{Tamii11} established the constraint 23 MeV $<L<$ 54
MeV from measurements of the dipole polarizability of $^{208}$Pb, and
a later study by Piekarewicz et al. \cite{Piekarewicz12}, who
attempted a more model-independent analysis, obtained 37 MeV $<L<$ 60
MeV (Figure~\ref{correl}).  A related experimental constraint
originates from measurements of the centroids of giant dipole
resonances (GDR) in spherical nuclei.  The hydrodynamical model of
Reference~\cite{LS89} showed that both the dipole polarizability and
the giant dipole resonance centroid energy are closely connected to
liquid droplet parameters.  The centroid energy in this model is
\begin{equation}\label{GDR}
E_{-1}\simeq\sqrt{{3\hbar^2\over m<r^2>}S_v\left(1+{5\over3}{S_s\over S_v}A^{-1/3}\right)^{-1}(1+\kappa)},
\end{equation}
where $\kappa$ is an enhancement factor arising from the velocity
dependence of the interaction and $<r^2>$ is the mean-square radius.
The factor $m<r^2>/(1+\kappa)$ does not significantly vary among
interactions for a given nucleus.  By using various Skyrme functions
to fit the centroid energy in $^{208}$Pb, Trippa et
al.~\cite{Trippa08} showed that Equation \ref{GDR} is approximately
equivalent to evaluating $S(u)$ at the subnuclear
density $0.1{\rm~fm}^{-3}$, leading to a small range of 23.3 MeV $<S_{0.1}<$ 24.9
MeV.  Although the relation among $S_{0.1}, S_v$ and $L$ is model
dependent, bounds \cite{Lim12} for a range of parameterizations are
indicated with the constraint band shown in Figure \ref{correl}.

A study of pygmy dipole resonances \cite{Carbone10} in $^{68}$Ni and
$^{132}$Sn imply that 31 MeV $<S_v<$ 33.6 MeV and 49.1 MeV $<L<$ 80
MeV.  However, Daoutidis \& Goriely \cite{DG11} claim that both
theoretical and experimental uncertainties pygmy dipole resonances
from being effective constraints, which is supported by the results
from Reinhard \& Nazarewicz~\cite{RN10} that show a lack of
correlation with $L$.  This constraint is not shown in Figure
\ref{correl}.

Danielewicz \& Lee~\cite{DL09} proposed isobaric analog states as
constraints on symmetry parameters, leading to $S_v\simeq32.9$ MeV and
$S_s\simeq96$ MeV, which is equivalent to $L\simeq113$ MeV.  However,
the data suggest that there could be significant curvature
contribution; refitting with curvature (unpublished) yields
$S_v\sim30$ MeV and $S_s\sim45$ MeV, which are equivalent to $L\sim50$
MeV, but with considerable errors. A more detailed analysis will be
necessary to produce better constraints.  Further constraints on $L$
alone were suggested from multifragmentation in intermediate-energy
heavy-ion collisions \cite{Shetty07}, leading to 40 MeV $<L<$ 125 MeV.
Although consistent with other constraints, neither result is useful
for further restricting parameter space.

Most of the experimental studies discussed above have a remarkable
consistency. The white overlap region shows the common predictions of
the experimental studies highlighted in Figure~\ref{correl}. (Note
that the finite-range droplet model would also predict a confidence
ellipse similar in extent and orientation to that of Reference
\cite{Kortelainen10}, but this was not calculated.)  It is remarkable
that the overlap is fully consistent with the observational inferences
made in Reference \cite{SLB10}.

\subsection{Neutron Matter and High-Density Constraints}
The symmetry parameters are also related to properties of pure neutron
matter at the saturation density.  If quartic and higher terms in the
symmetry energy can be neglected, the energy and pressure of pure
neutron matter at $n_s$ are given by Equations~(\ref{eose}) and
(\ref{eosp}) as $e_{Ns}=S_v-B$ and $p_{Ns}=Ln_s/3$, respectively.
However, depending on the assumed nuclear interaction, this might not
be a good approximation.  Considering only the non-relativistic,
non-interacting kinetic energy contributions, the errors made using
the quadratic approximation are
\begin{eqnarray}\label{diff}
e_{Ns,k}-S_k+B&=&\left({3\over5}2^{2/3}-{1\over3}\right){\hbar^2\over2m_b}\left({3\pi^2n_s\over2}\right)^{2/3}\equiv Q\simeq0.72{\rm~MeV},\cr
{p_{n,k}\over n_s}-{L_k\over3}&=&{2\over3}Q\simeq0.96{\rm~MeV}.
\end{eqnarray}
Although these errors are not large, contributions from the effective
mass and potential energy can be considerably larger, and the errors
amplify at high densities \cite{Steiner06}.  Therefore, care should be
taken in relating $S_v$ and $L$ to neutron matter properties.

Two recent independent studies \cite{HLPS10,GCR12} have estimated
properties of neutron matter based on chiral Lagrangian and quantum
Monte Carlo techniques with realistic two- and three-nucleon
interactions.  Employing realistic ranges of the magnitudes of these
interactions, and neglecting quartic and higher symmetry
contributions, the neutron matter studies lead to the symmetry parameter ranges shown in
Figure~\ref{correl}.  The close agreement between neutron matter
studies and laboratory constraints is encouraging, although the caveat
regarding the neglect of quartic and higher terms in the symmetry
expansion should be kept in mind.  However, given that neutron star
matter and pure neutron matter differ by relatively minor amounts
because $x<<1$, comparisons between the neutron matter and
astrophysics results do not have the ambiguity inherent in the
comparisons with laboratory studies.  In fact, the agreement with astrophysical
data is very good, not only for the derived symmetry parameters, but
also for the overall pressure-density relation at moderate densities
(Figure \ref{prhohsslb}).

\begin{figure}[h]
\includegraphics[angle=0,width=10cm]{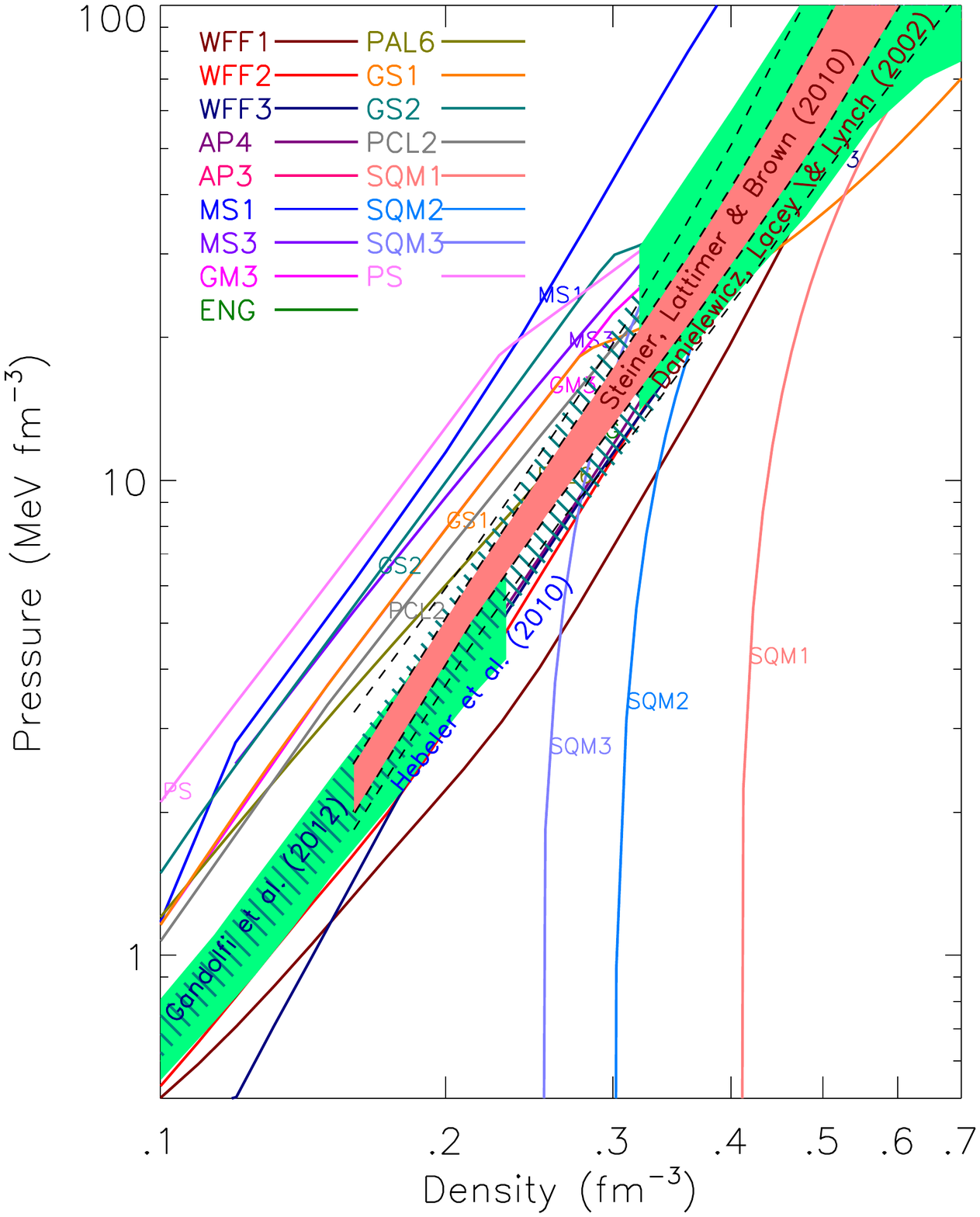}
\caption{Representative hadronic and strange quark matter (SQM) equations of state, as in Figure~\ref{prho} but with astrophysical 
results \cite{SLB10}, neutron matter calculations \cite{HLPS10,GCR12}, and heavy-ion studies \cite{DLL02} overlaid.}
\label{prhohsslb}
\end{figure}

There is a paucity of experimental information concerning the EOS at
densities modestly exceeding $n_s$.  Danielewicz et al.~\cite{DLL02}
analyzed the elliptic and transverse flows of matter in heavy-ion
(Au-Au) collisions to estimate the pressure as a function of density
in the range of 2 to 5 $n_s$.  However, the experimental conditions
involve large excitation energies and nearly symmetric matter. The
inferred pressures and energy densities must therefore be extrapolated
to zero temperature and to large neutron excesses that are
characteristic of neutron star matter.  Doing so introduces
considerable systematic errors.  Nevertheless, the results
(Figure~\ref{prhohsslb}) also compare favorably with the astrophysical
constraints.

\section{CONCLUSIONS}
An ever-increasing set of astrophysical data concerning neutron star
masses and radii has placed severe constraints on the nuclear EOS.
The available data suggests that neutron stars of canonical mass 1.4
M$_\odot$ have radii near 11.5 km, which strongly implies that the EOS
near the nuclear saturation density is relatively soft---in other
words, that the pressure there is relatively low.  At the same time,
the EOS must stiffen at densities exceeding $2-3n_s$ to permit the
existence of 2-M$_\odot$ neutron stars.  Although the existence of
exotic matter such as deconfined quarks, kaon condensates, and
hyperons, in neutron star interiors are not excluded by these
considerations, their role in determining the pressure-density
relation must not be great.  There are hints that even more massive
neutron star masses are waiting to be discovered.  The detection of
neutron stars with approximately 2.4 M$_\odot$ would be exciting
because of the potential to rule out the existence of some types of
exotic matter.  Even if the existence of such massive neutron stars
is not confirmed, the presence of exotic matter might yet be revealed
by observations of cooling neutron stars \cite{Page09}.  

The detection of gravitational radiation from mergers involving
neutron stars will present additional opportunities.  With an expected
detection rate of 0.4 to 40 per year with Advanced LIGO
\cite{Abadie10}, many new mass and radius estimates will become
available.  Given the observed narrow mass range of double--neutron
star binaries, observations of the peak frequency may additionally yield radius
measurements with an accuracy of 0.1 km \cite{Bauswein12a}.  Furthermore,
if the events are nearby enough, observations of postmerger
gravitational waves could set interesting upper limits to the neutron
star maximum mass.

The conclusions regarding the mass-radius relation and the nuclear
symmetry energy from astrophysical observations of quiescent neutron
stars in globular clusters and from Type I X-ray bursts are strongly
supported by the convergence in the
predicted properties of dense matter revealed by nuclear experiments
and theoretical neutron matter calculations.  Although the symmetry
parameters $S_v$ and $L$ do not completely describe the extrapolation
from symmetric matter to neutron-rich matter, it is gratifying that
their permissible ranges are now reasonably restricted, as shown in
Figure~\ref{correl}: 29.0 MeV $<S_v<$ 32.5 MeV and 41 MeV $<L<$ 60
MeV.  Continuing astrophysical observations combined with experimental
studies of giant dipole resonances and dipole polarizabilities, as
well as neutron skins, have great potential to further improve these
constraints.

\section*{Disclosure Statement}
The author is not aware of any affiliations, memberships, funding, or financial holdings that might be perceived as affecting the objectivity of this review.
\section*{Acknowledgments}

It is a pleasure to acknowledge my collaborators, including Ed Brown,
Kai Hebeler, Chris Pethick, Madappa Prakash, Achim Schwenk, Andrew
Steiner and Yeunhwan Lim. I also thank Rob Ferdman and David Nice for
providing details about pulsars.  Partial funding for this work was
generously provided by the US Department of Energy under grant
DE-FG02-87ER40317.



\end{document}